\documentclass{article}

\usepackage[T1]{fontenc}
\usepackage{geometry}
\usepackage{color}
\usepackage[english]{babel}
\usepackage{pifont}
\usepackage{mathrsfs}
\usepackage{amsmath}
\usepackage{amssymb}
\usepackage{graphicx}
\usepackage{setspace}
\usepackage[numbers,sort&compress]{natbib}
\usepackage[unicode=true,pdfusetitle,
 bookmarks=true,bookmarksnumbered=false,bookmarksopen=false,
 breaklinks=false,pdfborder={0 0 0},pdfborderstyle={},backref=false,colorlinks=true]
 {hyperref}
\hypersetup{
 urlcolor=blue, citecolor=blue}

\makeatletter

\newcommand{\noun}[1]{\textsc{#1}}

\newcommand{\lyxaddress}[1]{
	\par {\raggedright #1
	\vspace{1.4em}
	\noindent\par}
}

\usepackage{old-arrows}

\AtBeginDocument{
  
}

\makeatother

\begin{document}
\title{The correlation production in thermodynamics }
\author{Sheng-Wen Li}
\maketitle

\lyxaddress{Center for Quantum Technology Research, School of Physics, Beijing
Institute of Technology, Beijing 100081, China \\
Texas A\&M University, College Station, TX 77843}
\begin{abstract}
Macroscopic many-body systems always exhibit irreversible behaviors.
For example, the gas always tend to fill up the unoccupied area until
reaching the new uniform distribution, together with the irreversible
entropy increase. However, in principle, the underlying microscopic
dynamics of the many-body system, either the (quantum) von Neumann
or (classical) Liouville equation, guarantees the entropy of an isolated
system does not change with time, which is quite confusing comparing
with the macroscopic irreversibility. Notice that, due to the restrictions
in practical measurements, usually it is the partial information (e.g.,
marginal distribution, few-body observable expectation) that is directly
accessible to our observations, rather than the full ensemble state.
But indeed such partial information is sufficient to give most macroscopic
thermodynamic quantities, and they exhibits irreversible behaviors.
At the same time, there is some correlation entropy hiding in the
full ensemble, i.e., the mutual information between different marginal
distributions, but difficult to be sensed in practice. We notice that
such correlation production is indeed closely related to the macroscopic
entropy increase in the standard thermodynamics. In open systems,
the irreversible entropy production of the open system can be proved
to be equivalent with the correlation production between the open
system and its environment. During the free diffusion of an isolated
ideal gas, the correlation between the spatial and momentum distributions
is increasing monotonically, and it could well reproduce the entropy
increase result in the standard thermodynamics. In the presence of
particle collisions, the single-particle distribution always approaches
the Maxwell-Boltzmann distribution as its steady state, and its entropy
increase indeed indicates the correlation production between the particles.
In all these examples, the total entropy of the whole isolated system
keeps constant. In this sense, the macroscopic irreversibility and
the reversible microscopic dynamics coincide with each other. 
\end{abstract}

\section{Introduction }

\begin{figure}
\begin{centering}
\includegraphics[width=0.55\textwidth]{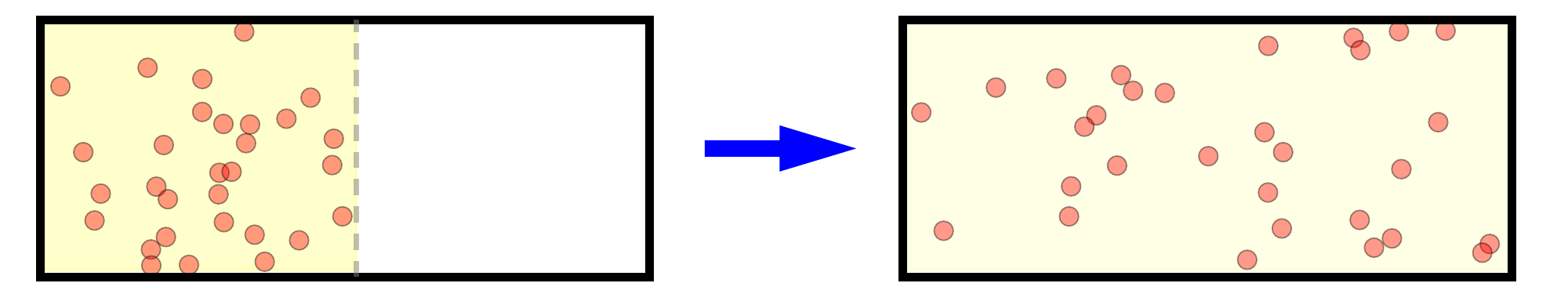}
\par\end{centering}
\caption{Demonstration for gas diffusion.}

\label{fig-expansion}
\end{figure}
Considering an isolated ideal gas initially occupying only part of
a box, after long enough time diffusion, the gas spreads all over
the volume uniformly (Fig.\,\ref{fig-expansion}). From the standard
macroscopic thermodynamics it is simple to show the gas entropy is
increased by $\Delta S=Nk_{\text{\textsc{b}}}\ln(V/V_{0})$, where
$V$ ($V_{0}$) is the final (initial) occupied volume \citep{huang_statistical_1987,le_bellac_equilibrium_2004}.

This is a quite typical example of the entropy increase in the macroscopic
thermodynamics. However, notice that isolated quantum systems always
follow the unitary evolution (described by the von Neumann equation
$\partial_{t}\hat{\rho}=i[\hat{\rho},\hat{{\cal H}}]$), and that
guarantees the von Neumann entropy $S_{\text{\textsc{v}}}[\hat{\rho}]=-\mathrm{tr}[\hat{\rho}\ln\hat{\rho}]$
does not change with time. In principle, this result should also apply
for many-body systems, then it seems inconsistent with the above entropy
increase in the standard macroscopic thermodynamics. 

Indeed, this is not a problem that only appears in quantum physics,
and classical physics has the same situation. For an isolated classical
system, the ensemble evolution follows the Liouville equation \citep{gibbs_elementary_1902,hobson_concepts_1971,huang_statistical_1987},
\begin{equation}
\partial_{t}\rho(\vec{P},\vec{Q},t)=-\{\rho(\vec{P},\vec{Q},t),\,{\cal H}\},
\end{equation}
which is derived from the Hamiltonian dynamics. Here $\{\cdot\,,\cdot\}$
is the Poisson bracket, and $\rho(\vec{P},\vec{Q},t)$ is the probability
density around the microstate $(\vec{P},\vec{Q}):=(\vec{p}_{1},\vec{p}_{2},\dots;\,\vec{q}_{1},\vec{q}_{2},\dots)$
at time $t$. As a result, the Gibbs entropy of the whole system keeps
a constant and never changes with time, 
\begin{equation}
\frac{d}{dt}S_{\text{\textsc{g}}}[\rho(\vec{P},\vec{Q},t)]=\frac{d}{dt}\Big[-\int d^{3N}p\,d^{3N}q\,\rho\ln\rho\Big]=0.
\end{equation}
Therefore, this constant entropy result exists in both quantum and
classical physics. 

This is rather confusing when comparing with our intuition of the
``irreversibility''\footnote{ In thermodynamics, a ``reversible
(irreversible)'' process means the system is (not) always in the
thermal equilibrium state at every moment. Throughout this paper,
we adopt the meaning in dynamics: for any initial condition, some
function (distribution, state, etc.) always approaches the same steady
state, then such kind of behavior is regarded as ``irreversible''.}
happening in the macroscopic world \citep{jaynes_gibbs_1965,hobson_concepts_1971,prigogine_time_1978,garrett_macroirreversibility_1991,uffink_compendium_2006,swendsen_explaining_2008,han_entropy_2015,dong_how_2017}.
Moreover, even if the particles have complicated nonlinear interactions,
although the system dynamics could be highly chaotic and unpredictable,
the (classical) Liouville or (quantum) unitary dynamics still guarantees
the entropy of isolated systems does not change with time.

On the other hand, if there is no inter-particle interaction, the
microstate evolution is well predictable, but the above irreversible
diffusion process could still happen until the gas refills the whole
volume uniformly. From this sense, it seems that the above contradiction
between the constant entropy and the appearance of macroscopic irreversibility
does not depend on whether there exist complicated interactions. We
need to ask: how could the macroscopic irreversibility and entropy
increase arise from the underlying microscopic dynamics, which is
reversible with time-reversal symmetry \citep{mackey_dynamic_1989,prigogine_time_1978}?

Recently, it is noticed that the irreversible entropy production in
open systems indeed is deeply related with correlation between the
open system and its environment \citep{esposito_entropy_2010,manzano_entropy_2016,alipour_correlations_2016,li_production_2017}.
In an open system, the entropy of the system itself can either increase
or decrease, depending on whether it is absorbing or emitting heat
to its environment. Subtracting such thermal entropy due to the heat
exchange, the rest part of the system entropy change is called the
irreversible entropy production \citep{bergmann_new_1955,de_groot_non-equilibrium_1962,nicolis_self-organization_1977,reichl_modern_2009,kondepudi_modern_2014},
and that increases monotonically with time until reaching the thermal
equilibrium. 

Under proper approximations, we can prove indeed the thermal entropy
change due to the heat exchange is just equal to the entropy change
of the environment state \citep{aurell_von_2015,li_production_2017,you_entropy_2018},
and the irreversible entropy production is equivalent as the correlation
generation between the open system and its environment, which is measured
by the relative entropy \citep{esposito_entropy_2010,pucci_entropy_2013,manzano_entropy_2016,strasberg_quantum_2017,manzano_quantum_2018}
or their mutual information \citep{li_production_2017,you_entropy_2018}.
At the same time, the system and its environment together as a whole
system maintains constant entropy during the evolution. In this sense,
the constant global entropy and the increase of correlation well coincide
with each other. Moreover, when the baths are non-thermal states,
which are beyond the application scope of the standard macroscopic
thermodynamics, we could see such correlation production still applies
(see Sec.\,2.4).

That is to say, due to the practical restrictions of measurements,
indeed some correlation information hiding in the global state is
difficult to be sensed, and that results to the appearance of the
macroscopic irreversibility as well as the entropy increase. In principle,
such correlation understanding could also apply for isolated systems.
Indeed, in the above diffusion example, our observation that ``the
gas spread all over the volume uniformly'' is implicitly focused
on the spatial distribution only, rather than the total ensemble state. 

For the classical ideal gas with no inter-particle interactions, the
Liouville equation for the ensemble evolution can be exactly solved
\citep{hobson_irreversibility_1966,hobson_concepts_1971,swendsen_explaining_2008}.
Notice that in practice, it is the spatial and momentum distributions
that are directly measured, but not the full ensemble state. We can
prove that the spatial distribution $\mathscr{P}_{\mathtt{x}}(x,t)$,
as a marginal distribution of the whole ensemble, always approaches
the new uniform one as its steady state. Moreover, by examining the
correlation between the spatial and momentum distributions, we can
see their correlation increases monotonically, and could reproduce
the entropy increase in the standard thermodynamics. At the same time,
the total ensemble state $\rho(\vec{P},\vec{Q},t)$ keeps constant
entropy during the diffusion process (see Sec.\,3).

For the non-ideal gas with weak particle interactions, the dynamics
of the single-particle probability distribution function (PDF) $f(\mathbf{p},\mathbf{r},t)$
can be described by the Boltzmann equation \citep{huang_statistical_1987,uffink_compendium_2006}.
According to the Boltzmann \emph{H}-theorem, $f(\mathbf{p},\mathbf{r},t)$
always approaches the Maxwell-Boltzmann (MB) distribution as its steady
state, and its entropy increases monotonically. Notice that the single-particle
PDF $f(\mathbf{p},\mathbf{r},t)$ is a marginal distribution of the
full ensemble state $\rho(\vec{P},\vec{Q},t)$, which is obtained
by averaging out all the other particles. Thus $f(\mathbf{p},\mathbf{r},t)$
does not contain the particle correlations, and the increase of its
entropy indeed implicitly reflects the increase of the inter-particle
correlations, which exactly reproduces the entropy increase result
in the standard macroscopic thermodynamics. At the same time, the
total ensemble $\rho(\vec{P},\vec{Q},t)$ still follows the Liouville
equation with constant entropy.

The correlation production between the particles could also help us
understand the Loschmidt paradox: when we consider the ``backward''
evolution, since significant particle correlations have established
\citep{chliamovitch_kinetic_2017}, the molecular-disorder assumption,
which is the most crucial approximation in deriving the Boltzmann
equation, indeed does not hold. Therefore, the Boltzmann equation
as well as the \emph{H}-theorem of entropy increase does not apply
for the ``backward'' evolution (see Sec.\,4).

In sum, the global state keeps constant entropy, but in practice,
usually it is the partial information (e.g., marginal distribution,
single-particle observable expectations) that is directly accessible
to our observation, and that gives rise to the appearance of the macroscopic
irreversibility \citep{li_production_2017,you_entropy_2018,zhang_general_2008,zhang_information_2009,esposito_entropy_2010,pucci_entropy_2013,horowitz_equivalent_2014,alipour_correlations_2016,manzano_entropy_2016,strasberg_quantum_2017,kalogeropoulos_time_2018,hobson_irreversibility_1966,swendsen_explaining_2008,cramer_exact_2008,eisert_quantum_2015}.
The entropy increase in the standard macroscopic thermodynamics indeed
reflects the correlation increase between different degrees of freedom
(DoF) in the many-body system. In this sense, the reversibility of
microscopic dynamics (for the global state) and the macroscopic irreversibility
(for the partial information) coincide with each other. More importantly,
this correlation understanding applies for both quantum and classical
systems, and for both open and isolated systems; besides, it does
not depends on whether there exist complicated particle interactions,
and also can be used to describe time-dependent non-equilibrium systems.

\section{The correlation production in open systems}

In this section, we first discuss the thermodynamics of an open system,
which is surrounded by an environment exchanging energy with it. The
open system can absorb or emit heat to the environment, as a result,
the entropy of the open system itself can either increase or decrease.
Thus  the thermodynamic irreversibility is not simply related to the
entropy change of the open system alone, but should be described by
the ``\emph{irreversible entropy}'', which increases monotonically
with time.

Here we first give a brief review about this formalism for the irreversible
entropy production, which is an equivalent statement for the second
law. Then we will show indeed this irreversible entropy production
in open systems is just equivalent with the correlation increase between
the system and its environment \citep{zhang_general_2008,esposito_entropy_2010,li_production_2017},
which is measured by their mutual information. Moreover, if the baths
contacting with the system are not canonical thermal ones, the temperatures
are no longer well defined, and this situation is indeed not within
the applicable scope of the second law in the standard thermodynamics,
but we will see the the correlation production still applies in this
case.

\subsection{The irreversible entropy production rate \label{subsec:EPr}}

Now we first briefly review the formalism of entropy production \citep{bergmann_new_1955,de_groot_non-equilibrium_1962,nicolis_self-organization_1977,reichl_modern_2009,kondepudi_modern_2014}.
The entropy change $dS$ of an open system can be regarded as coming
from two origins, i.e., 
\begin{equation}
dS=dS_{\mathrm{e}}+dS_{\mathrm{i}}\label{eq:dSi-0}
\end{equation}
 where $dS_{\mathrm{e}}$ comes from the heat exchange with external
baths, and $dS_{\mathrm{i}}$ is regarded as the irreversible entropy
change. The exchanging part $dS_{\mathrm{e}}$ can be either positive
or negative, indicating the heat absorbing or emitting of the system.
But the irreversible entropy change $dS_{\mathrm{i}}$, as stated
by the second law, should always be positive.

If the system is contacted with a thermal bath in the equilibrium
state with temperature $T$, the entropy change due to the heat exchange
can be written as $dS_{\mathrm{e}}=\text{\text{\dj}}Q/T$ (hereafter
we refer it as the \emph{thermal entropy}), where $\text{\dj}Q$ is
the heat absorbed by the system. Then the second law can be expressed
as
\begin{equation}
dS_{\mathrm{i}}=dS-\frac{\text{\text{\text{\dj}}}Q}{T}\ge0,
\end{equation}
where the equality holds only for reversible processes. This is just
the Clausius inequality for an infinitesimal process \citep{kondepudi_modern_2014,de_groot_non-equilibrium_1962,huang_statistical_1987}.

\begin{figure}
\begin{centering}
\includegraphics[width=0.3\textwidth]{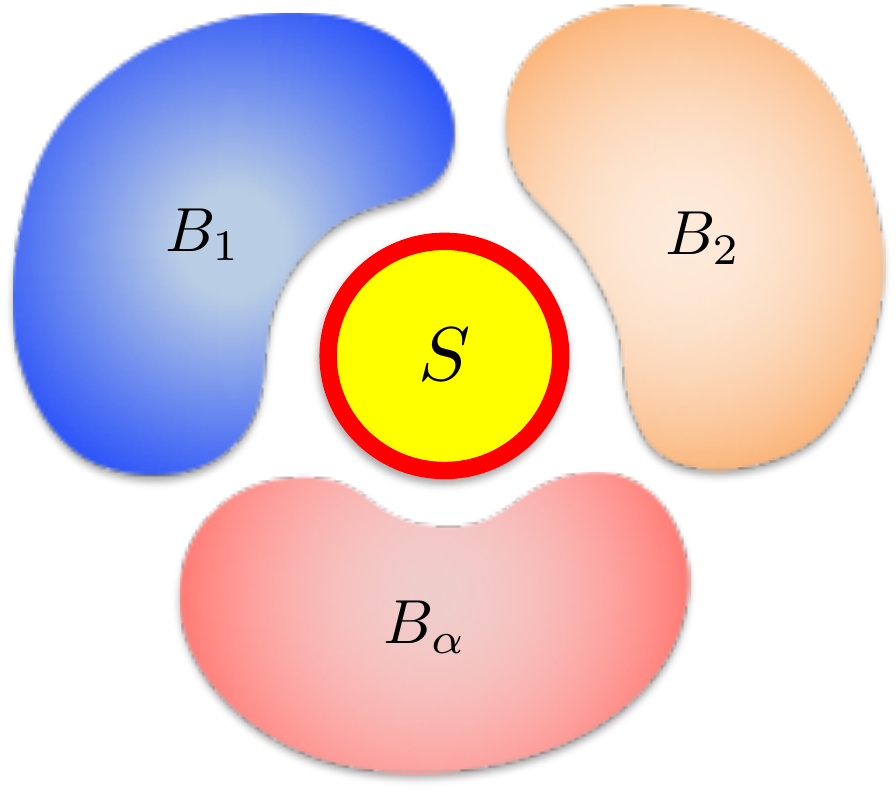}
\par\end{centering}
\caption{Demonstration for an open system $S$ surrounded by several independent
baths $B_{\alpha}$.}

\label{fig-mBaths}
\end{figure}
More generally, if the system contacts with multiple independent thermal
baths with different temperatures $T_{\alpha}$ at the same time (Fig.\,\ref{fig-mBaths}),
the irreversible entropy change should be generalized as \citep{bergmann_new_1955}
\begin{equation}
dS_{\mathrm{i}}=dS-\sum_{\alpha}\frac{\text{\text{\text{\dj}}}Q_{\alpha}}{T_{\alpha}}\ge0,\label{eq:dSi}
\end{equation}
where $\text{\text{\text{\dj}}}Q_{\alpha}$ is the heat absorbed from
bath-$\alpha$ \citep{de_groot_non-equilibrium_1962,kondepudi_modern_2014}.
For example, for a system contacting with two thermal baths with temperatures
$T_{1,2}$, in the steady state, we have $dS=0$ and $\text{\text{\text{\dj}}}Q_{1}=-\text{\text{\text{\dj}}}Q_{2}$,
thus the above equation gives \citep{kondepudi_modern_2014} 
\begin{equation}
-\text{\text{\text{\dj}}}Q_{1}\big(\frac{1}{T_{1}}-\frac{1}{T_{2}}\big)\ge0.\label{eq:T1T2}
\end{equation}
It is easy to verify $\text{\text{\text{\dj}}}Q_{1}=-\text{\text{\text{\dj}}}Q_{2}>0$
always comes together with $T_{1}>T_{2}$, and vice versa. That means,
the heat always flows from the high temperature area to the low temperature
area, which is just the Clausius statement of the second law.

Therefore, for an open system, the second law can be equivalently
expressed as a simple inequality $dS_{\mathrm{i}}\ge0$, which means
the irreversible entropy change always increases monotonically. This
can be also expressed by the \emph{entropy production rate} (EPr),
which is defined as 
\begin{equation}
R_{\text{\textsc{ep}}}:=\frac{dS_{\mathrm{i}}}{dt}=\frac{dS}{dt}-\sum_{\alpha}\frac{1}{T_{\alpha}}\frac{dQ_{\alpha}}{dt},\label{eq:R-EPr}
\end{equation}
and $R_{\text{\textsc{ep}}}\ge0$ is equivalent as saying the irreversible
entropy keeps increasing.

Besides the equivalence with the standard second law statements, the
entropy production formalism also provides a proper way to quantitively
study the non-equilibrium thermodynamics. Considering there is only
one thermal bath, the system would get thermal equilibrium with the
bath in the steady state. At this time, the system state no longer
changes, and there is no net heat exchange between the system and
the bath, thus $R_{\text{\textsc{ep}}}\rightarrow0$ when $t\rightarrow\infty$. 

In contrast, if the system contacts with multiple thermal baths with
different temperatures, in the steady state, although the system state
no longer changes with time, there still exists net heat flux between
the system and the baths. Therefore, different from the thermal equilibrium,
such a steady state is a \emph{stationary non-equilibrium state} \citep{bergmann_new_1955}.
Notice that in this case the EPr remains a finite positive value $R_{\text{\textsc{ep}}}>0$
{[}see the example of Eq.\,(\ref{eq:T1T2}){]} when $t\rightarrow\infty$,
which indicates there is still on-going production of irreversible
entropy. Therefore, $R_{\text{\textsc{ep}}}=0$ (or $>0$) well indicates
whether (or not) the system is in the thermal equilibrium state.

Here the above discussions about the entropy production apply for
both classical and quantum systems, as long as the quantities like
$\dot{S}$ and $\dot{Q}_{\alpha}$ are calculated by the classical
ensemble or quantum state correspondingly. 

\subsection{The production rate of the system-bath correlation \label{subsec:S-B-correlation}}

Now we will show the above EPr is indeed equivalent as the production
rate of the correlation between the open system and its environment.
Usually the dynamics of the open system alone is more often concerned
in literature. The baths, due to their large size, are usually considered
as unaffected by the system, and only provides a background with fluctuations.
But the system surely has influence to its environment \citep{li_production_2017,you_entropy_2018,wang_magnetic_2018}.
For example, when the system emits energy, this energy is indeed added
to the environment. To study the correlation between the system and
its environment, here we also need to know the dynamics of the whole
environment.

\textbf{\vspace{0.2em}\\ \noindent Quantum case:} Here we first consider
a quantum system contacting with several independent thermal baths
with temperatures $T_{\alpha}$. Initially, each bath-$\alpha$ stays
in the canonical thermal state 
\begin{equation}
\hat{\rho}_{\text{\textsc{b}},\alpha}(0)=\frac{1}{Z_{\alpha}}\exp[-\frac{\hat{H}_{\text{\textsc{b}},\alpha}}{T_{\alpha}}],
\end{equation}
with $Z_{\alpha}$ as the normalization factor. The exact changing
rate of the information entropy of bath-$\alpha$ is given by $\frac{d}{dt}S_{\text{\textsc{b}},\alpha}(t)=-\mathrm{tr}[\dot{\hat{\rho}}_{\text{\textsc{b}},\alpha}(t)\ln\hat{\rho}_{\text{\textsc{b}},\alpha}(t)]$
\footnote{Formally, the von Neumann entropy $S_{\text{\textsc{v}}}[\hat{\rho}]=-\mathrm{tr}[\hat{\rho}\ln\hat{\rho}]$
gives $\dot{S}_{\text{\textsc{v}}}=-\mathrm{tr}\big[\dot{\hat{\rho}}\cdot\ln\hat{\rho}+\hat{\rho}\cdot(\hat{\rho}^{-1}\cdot\dot{\hat{\rho}})\big]=-\mathrm{tr}[\dot{\hat{\rho}}\ln\hat{\rho}]$.}.
To make further calculation, we assume the bath state $\hat{\rho}_{\text{\textsc{b}},\alpha}(t)$
does not change too much from the initial state, thus $\ln\hat{\rho}_{\text{\textsc{b}},\alpha}(t)=\ln[\hat{\rho}_{\text{\textsc{b}},\alpha}(0)+\delta\hat{\rho}_{t}]\simeq\ln\hat{\rho}_{\text{\textsc{b}},\alpha}(0)+o(\delta\hat{\rho}_{t})$,
then the bath entropy change $\dot{S}_{\text{\textsc{b}},\alpha}(t)$
becomes \citep{aurell_von_2015,li_production_2017,you_entropy_2018,manzano_quantum_2018}
\begin{align}
\dot{S}_{\text{\textsc{b}},\alpha}(t) & \simeq-\mathrm{tr}[\dot{\hat{\rho}}_{\text{\textsc{b}},\alpha}(t)\ln\hat{\rho}_{\text{\textsc{b}},\alpha}(0)]\nonumber \\
 & =-\mathrm{tr}\Big\{\dot{\hat{\rho}}_{\text{\textsc{b}},\alpha}(t)\cdot\ln\big(\frac{1}{Z_{\alpha}}\exp[-\frac{\hat{H}_{\text{\textsc{b}},\alpha}}{T_{\alpha}}]\big)\Big\}=\frac{1}{T_{\alpha}}\frac{d}{dt}\langle\hat{H}_{\text{\textsc{b}},\alpha}\rangle.\label{eq:S_Ba}
\end{align}

Notice that here $\frac{d}{dt}\langle\hat{H}_{\text{\textsc{b}},\alpha}\rangle$
is the energy increase of bath-$\alpha$, thus it is just equal to
the energy loss of the system to bath-$\alpha$ (i.e., $-\dot{Q}_{\alpha}$)
when the system-bath interaction strength is negligibly small. Therefore,
the above EPr $R_{\text{\textsc{ep}}}$ {[}Eq.\,(\ref{eq:R-EPr}){]}
can be rewritten as $R_{\text{\textsc{ep}}}\simeq\dot{S}_{\text{\textsc{s}}}(t)+\sum_{\alpha}\dot{S}_{\text{\textsc{b}},\alpha}(t)$.

Since initially the different baths are independent from each other
and do not interact with each other directly, we assume they cannot
generate significant correlations during the evolution, thus the entropy
of the whole environment is simply the summation of that from each
single bath, namely, $S_{\text{\textsc{b}}}(t)\simeq\sum_{\alpha}S_{\text{\textsc{b}},\alpha}(t)$.
Therefore, the above EPr $R_{\text{\textsc{ep}}}$ can be further
rewritten as
\begin{equation}
R_{\text{\textsc{ep}}}\simeq\dot{S}_{\text{\textsc{s}}}(t)+\dot{S}_{\text{\textsc{b}}}(t)=\frac{d}{dt}[S_{\text{\textsc{s}}}+S_{\text{\textsc{b}}}-S_{\text{\textsc{sb}}}]=\frac{d}{dt}{\cal I}_{\text{\textsc{sb}}}(t).\label{eq:R_I_SB}
\end{equation}
The above equality holds because the whole \noun{s}+\noun{b} system
is an isolated system and follows the unitary evolution, thus the
von Neumann entropy of the whole \noun{s+b} state does not change
with time, namely, $\frac{d}{dt}S_{\text{\textsc{sb}}}:=\dot{S}_{\text{\textsc{v}}}[\hat{\rho}_{\text{\textsc{sb}}}(t)]=0$
\citep{nielsen_quantum_2000}.

Therefore, the production rate of the irreversible entropy $R_{\text{\textsc{ep}}}$
is just equivalent with the production rate of the mutual information
between the system and its environment, ${\cal I}_{\text{\textsc{sb}}}=S_{\text{\textsc{s}}}+S_{\text{\textsc{b}}}-S_{\text{\textsc{sb}}}$,
which measures their correlation \citep{nielsen_quantum_2000}. That
means, the second law statement that the irreversible entropy keeps
increasing ($R_{\text{\textsc{ep}}}\ge0$) can be also equivalently
understood as, the correlation between the system and its environment,
as measured by the mutual information, always keeps increasing until
they get the equilibrium.

\textbf{\vspace{0.2em}\\ \noindent Classical case:} The above discussions
about quantum open systems also applies for classical ones. For classical
systems, the initial state of bath-$\alpha$ should be represented
by the canonical ensemble distribution 
\begin{equation}
\rho_{\text{\textsc{b}},\alpha}(\vec{P},\vec{Q},t=0)=\frac{1}{Z_{\alpha}}\exp[-\frac{1}{T_{\alpha}}H_{\text{\textsc{b}},\alpha}(\vec{P},\vec{Q})],
\end{equation}
where $(\vec{P},\vec{Q}):=(\vec{p}_{1},\vec{p}_{2},\dots;\vec{q}_{1},\vec{q}_{2},\dots)$
denotes the momentums and positions of the DoF in bath-$\alpha$.
Then we consider the changing rate of the Gibbs entropy of bath-$\alpha$,
and that is\footnote{Due to the integration over $\vec{P}$ and $\vec{Q}$,
the functional $S_{\text{G}}\big[\rho_{\text{\textsc{b}},\alpha}(\vec{P},\vec{Q},t)\big]$
does not explicitly contain $\vec{P}$ and $\vec{Q}$, thus $\frac{d}{dt}S_{\text{G}}=\big[\partial_{t}+\dot{\vec{P}}\cdot\partial_{\vec{P}}+\dot{\vec{Q}}\cdot\partial_{\vec{Q}}\big]S_{\text{G}}=\partial_{t}S_{\text{G}}$.
This also applies for the average expectations like $\big\langle H_{\text{\textsc{b}},\alpha}(\vec{P},\vec{Q})\big\rangle$.}
\begin{align}
\frac{d}{dt}S_{\text{\textsc{g}}}\big[\rho_{\text{\textsc{b}},\alpha}(\vec{P},\vec{Q},t)\big] & =-\int d^{3N}p\,d^{3N}q\,\partial_{t}\rho_{\text{\textsc{b}},\alpha}(t)\ln\rho_{\text{\textsc{b}},\alpha}(t)\simeq-\int d^{3N}p\,d^{3N}q\,\partial_{t}\rho_{\text{\textsc{b}},\alpha}(t)\ln\rho_{\text{\textsc{b}},\alpha}(0)\nonumber \\
 & =-\int d^{3N}p\,d^{3N}q\,\partial_{t}\rho_{\text{\textsc{b}},\alpha}(t)\cdot\big[-\frac{1}{T_{\alpha}}H_{\text{\textsc{b}},\alpha}(\vec{P},\vec{Q})\big]=\frac{1}{T_{\alpha}}\frac{d}{dt}\big\langle H_{\text{\textsc{b}},\alpha}(\vec{P},\vec{Q})\big\rangle.
\end{align}
Here we adopted the similar approximation $\ln\rho_{\text{\textsc{b}},\alpha}(\vec{P},\vec{Q},t)\simeq\ln\rho_{\text{\textsc{b}},\alpha}(\vec{P},\vec{Q},0)$
as above, and this result is simply the classical counterpart of Eq.\,(\ref{eq:S_Ba}). 

Therefore, for classical open systems, the EPr $R_{\text{\textsc{ep}}}$
in Eq.\,(\ref{eq:R-EPr}) also can be rewritten as $R_{\text{\textsc{ep}}}=\frac{d}{dt}(S_{\text{\textsc{s}}}+S_{\text{\textsc{b}}})$.
Further, since the whole \noun{s+b }system is an isolated system,
its dynamics follows the Liouville equation, thus the Gibbs entropy
of the whole \noun{s+b }system does not change with time, i.e., $\frac{d}{dt}S_{\text{\textsc{sb}}}=0$.
Therefore, the equivalence between the irreversible entropy production
and the system-bath correlation {[}Eq.\,(\ref{eq:R_I_SB}){]} also
holds for classical systems. That means, for both classical and quantum
open systems contacting with thermal baths, the second law can be
equivalently stated as, the correlation between the system and its
environment, which is measured by their mutual information, always
keeps increasing.

\subsection{Master equation representation}

Besides the above general discussions, the time-dependent dynamics
of the open system, either classical or quantum, can be quantitively
described by a master equation. With the help of the master equations,
the above EPr can be further written in a more detailed form. Here
we will show this for both classical and quantum cases.

\textbf{\vspace{0.2em}\\ \noindent Classical case:} For a classical
open system, the interaction with the baths would lead to the probability
transition between its different states, and this dynamics is usually
described by the Pauli master equation \citep{gardiner_handbook_1985}
\begin{equation}
\dot{p}_{n}=\sum_{\alpha}\sum_{m}L_{n\leftarrow m}^{(\alpha)}\,p_{m}-L_{m\leftarrow n}^{(\alpha)}\,p_{n},
\end{equation}
which is a Markovian process. Here $p_{n}$ is the probability to
find the system in state-$n$ (whose energy is $\text{\textsc{e}}_{n}$),
and $L_{n\leftarrow m}^{(\alpha)}$ is the probability transition
rate from state-$m$ to state-$n$ due to the interaction with the
thermal bath-$\alpha$. The back and forth transition rates between
states-$m,n$ should satisfy the following ratio \citep{bergmann_new_1955,breuer_theory_2002}
\begin{equation}
\frac{L_{m\leftarrow n}^{(\alpha)}}{L_{n\leftarrow m}^{(\alpha)}}=\exp[-\frac{1}{T_{\alpha}}(\text{\textsc{e}}_{m}-\text{\textsc{e}}_{n})],\label{eq:jumpRatio}
\end{equation}
which means the ``downward'' transition to the low energy state
is faster than the ``upward'' one by a Boltzmann factor. In the
case of only one thermal bath, with this relation, the detailed balance
$L_{n\leftarrow m}^{(\alpha)}\,p_{m}-L_{m\leftarrow n}^{(\alpha)}\,p_{n}=0$
simply leads to the Boltzmann distribution $p_{n}:p_{m}=e^{-\text{\textsc{e}}_{n}/T}:e^{-\text{\textsc{e}}_{m}/T}$
in the steady state. 

If there are multiple thermal baths, the energy average $\langle E\rangle=\sum_{n}\text{\textsc{e}}_{n}p_{n}$
gives an energy-flow conservation relation
\begin{equation}
\partial_{t}\langle E\rangle=\sum_{\alpha}J_{\alpha},\qquad J_{\alpha}:=\sum_{m,n}(L_{n\leftarrow m}^{(\alpha)}\,p_{m}-L_{m\leftarrow n}^{(\alpha)}\,p_{n})\text{\textsc{e}}_{n},
\end{equation}
 thus $J_{\alpha}$ is the heat current flowing into the system from
bath-$\alpha$ ($\dot{Q}_{\alpha}$). We can put these relations,
as well as the Gibbs entropy of the system $S_{\text{\textsc{g}}}=-\sum_{n}p_{n}\ln p_{n}$,
into the above EPr (\ref{eq:R-EPr}), obtaining \citep{cai_entropy_2014}
\begin{align}
R_{\text{\textsc{ep}}} & =\sum_{\alpha}\sum_{m,n}-(L_{n\leftarrow m}^{(\alpha)}\,p_{m}-L_{m\leftarrow n}^{(\alpha)}\,p_{n})\ln p_{n}-\frac{\text{\textsc{e}}_{n}}{T_{\alpha}}(L_{n\leftarrow m}^{(\alpha)}\,p_{m}-L_{m\leftarrow n}^{(\alpha)}\,p_{n})\nonumber \\
 & =\sum_{\alpha}\sum_{m,n}\frac{1}{2}(L_{n\leftarrow m}^{(\alpha)}\,p_{m}-L_{m\leftarrow n}^{(\alpha)}\,p_{n})(\ln\frac{e^{-\text{\textsc{e}}_{n}/T_{\alpha}}}{p_{n}}-\ln\frac{e^{-\text{\textsc{e}}_{m}/T_{\alpha}}}{p_{m}})\nonumber \\
 & =\sum_{\alpha}\sum_{m,n}\frac{1}{2}(L_{n\leftarrow m}^{(\alpha)}\,p_{m}-L_{m\leftarrow n}^{(\alpha)}\,p_{n})\,\ln\big(L_{n\leftarrow m}^{(\alpha)}\,p_{m}/L_{m\leftarrow n}^{(\alpha)}\,p_{n}\big).\label{eq:R-classical}
\end{align}
Notice that each summation term must be non-negative, thus we always
have $R_{\text{\textsc{ep}}}\ge0$, which is just consistent with
the above second law statement that the irreversible entropy keeps
increasing. $R_{\text{\textsc{ep}}}=0$ holds only when $L_{n\leftarrow m}^{(\alpha)}\,p_{m}=L_{m\leftarrow n}^{(\alpha)}\,p_{n}$
for any $\alpha$, and this is possible only when all the baths have
the same temperature, which means the thermal equilibrium. Otherwise,
in the steady state, although it is time-independent, there still
exists non-equilibrium flux flowing across the system, and that is
indicated by $R_{\text{\textsc{ep}}}>0$.

\textbf{\vspace{0.2em}\\ \noindent Quantum case:} For a quantum system
weakly coupled with the multiple thermal baths, usually its dynamics
can be described by the GKSL (Lindblad) equation \citep{gorini_completely_1976,lindblad_generators_1976},
\begin{equation}
\dot{\hat{\rho}}=i[\hat{\rho},\hat{H}_{\text{\textsc{s}}}]+\sum_{\alpha}{\cal L}_{\alpha}[\hat{\rho}].\label{eq:QME}
\end{equation}
where $\hat{\rho}$ is the system state and ${\cal L}_{\alpha}[\hat{\rho}]$
describes the dissipation due to bath-$\alpha$. Using the von Neumann
entropy $S_{\text{\textsc{v}}}[\hat{\rho}]=-\mathrm{tr}[\hat{\rho}\ln\hat{\rho}]$
and heat current $\dot{Q}_{\alpha}=\mathrm{tr}\big[\hat{H}_{\text{\textsc{s}}}\cdot{\cal L}_{\alpha}[\hat{\rho}]\big]$,
the EPr (\ref{eq:R-EPr}) can be rewritten as the following Spohn
formula \citep{spohn_entropy_1978,spohn_irreversible_1978,alicki_quantum_1979,boukobza_three-level_2007,kosloff_quantum_2013,kosloff_quantum_2017,cai_entropy_2014}
\begin{align}
R_{\text{\textsc{ep}}} & =-\mathrm{tr}\big[\dot{\hat{\rho}}\ln\hat{\rho}\big]+\sum_{\alpha}\mathrm{tr}\big[{\cal L}_{\alpha}[\hat{\rho}]\cdot\ln\hat{\rho}_{\mathrm{ss}}^{(\alpha)}\big]\nonumber \\
 & =\sum_{\alpha}\mathrm{tr}\big[(\ln\hat{\rho}_{\mathrm{ss}}^{(\alpha)}-\ln\hat{\rho}){\cal L}_{\alpha}[\hat{\rho}]\big]:=R_{\mathrm{Sp}}.\label{eq:Spohn}
\end{align}
Here $\hat{\rho}_{\mathrm{ss}}^{(\alpha)}$ satisfies ${\cal L}_{\alpha}[\hat{\rho}_{\mathrm{ss}}^{(\alpha)}]=0$,
and we call it the \emph{partial steady state} associated with bath-$\alpha$.
If the system only interacts with bath-$\alpha$, then $\hat{\rho}_{\mathrm{ss}}^{(\alpha)}$
should be its steady state when $t\rightarrow\infty$. Clearly, $\hat{\rho}_{\mathrm{ss}}^{(\alpha)}$
should be the thermal state ($\sim\exp[-\hat{H}_{\text{\textsc{s}}}/T_{\alpha}]$)
when bath-$\alpha$ is the canonical thermal one with temperature
$T_{\alpha}$, and the term $\chi_{\alpha}:=\mathrm{tr}\big[{\cal L}_{\alpha}[\hat{\rho}]\cdot\ln\hat{\rho}_{\mathrm{ss}}^{(\alpha)}\big]=-\dot{Q}_{\alpha}/T_{\alpha}$
is the corresponding exchange of thermal entropy.

The positivity of $R_{\mathrm{Sp}}$ is not so obvious as the classical
case (\ref{eq:R-classical}), but still we can prove $R_{\mathrm{Sp}}\ge0$,
if the master equation (\ref{eq:QME}) has the standard GKSL form
(see the proof in Appendix of Ref.\,\citep{li_production_2017} or
Ref.\,\citep{spohn_entropy_1978,spohn_irreversible_1978}). The GKSL
form of the master equation (\ref{eq:QME}) indicates it describes
a Markovian process \citep{gorini_completely_1976,lindblad_generators_1976,li_non-markovianity_2016},
which is similar like the above classical case. Again this is consistent
with the above discussions about the second law statement. 

\textbf{\vspace{0.2em}\\ \noindent  Remark: }In the above discussions,
we all focused on the case that the baths are canonical thermal ones.
As a result, in the above master equations, the transition rate ratios
(\ref{eq:jumpRatio}) appear as the Boltzmann factors, and the partial
steady states $\hat{\rho}_{\mathrm{ss}}^{(\alpha)}$ of the system
are the canonical thermal states. Strictly speaking, only for canonical
thermal baths, the temperature $T$ is well defined, and the thermal
entropy $dS_{\mathrm{e}}=\text{\text{\text{\dj}}}Q/T$ can be applied,
as well as the above EPr (\ref{eq:R-EPr}), which is the starting
point to derive the master equation representations Eqs.\,(\ref{eq:R-classical},
\ref{eq:Spohn}).

If the baths are non-thermal states, there is no well-defined temperature,
thus the above EPr in the standard thermodynamics in Sec.\,\ref{subsec:EPr},
especially the thermal entropy $dS_{\mathrm{e}}=dQ/T$, does not apply.
But master equations still can be used to study the dynamics of such
systems. Due to the interaction with non-thermal baths, the transition
rate ratios (\ref{eq:jumpRatio}) do not need to be the Boltzmann
factors, but we can verify the last line of Eq.\,(\ref{eq:R-classical})
still remains positive. Thus Eq.\,(\ref{eq:R-classical}) can be
regarded as a generalized EPr beyond the standard thermodynamics,
however, now it is unclear to tell its physical meaning, as well as
its relation with the non-thermal bath.

The quantum case has the same situation. If the master equation (\ref{eq:QME})
comes from non-thermal baths, the partial steady state $\hat{\rho}_{\mathrm{ss}}^{(\alpha)}$
would not be the thermal state with the temperature of bath-$\alpha$,
but the Spohn formula {[}last line of Eq.\,(\ref{eq:Spohn}){]} still
remains positive \citep{li_production_2017,spohn_entropy_1978,spohn_irreversible_1978}.
However, in this case the physical meaning of the Spohn formula (\ref{eq:Spohn})
is not clear now. 

In the following example of an open quantum system interacting with
non-thermal baths, we will show that, although it is beyond the applicable
scope of the standard thermodynamics, the Spohn formula (\ref{eq:Spohn})
is still equal to the production rate of the system-bath correlation,
which is the same as the thermal bath case in Sec.\,\ref{subsec:S-B-correlation},
and the term $\chi_{\alpha}=\mathrm{tr}\big[{\cal L}_{\alpha}[\hat{\rho}]\cdot\ln\hat{\rho}_{\mathrm{ss}}^{(\alpha)}\big]$
is just equal to the informational entropy change of bath-$\alpha$.

\subsection{Contacting with squeezed thermal baths}

When the heat baths contacting with the system are not canonical thermal
ones, it is possible to construct a heat engine that ``seemingly''
works beyond the Carnot bound. For example, in an optical cavity,
a collection of atoms with non-vanishing quantum coherence can be
used to generate light force to do mechanical work by pushing the
cavity well \citep{scully_extracting_2003}; a squeezed light field
can be used to as the reservoir for an harmonic oscillator which expands
and compresses as a heat engine \citep{rosnagel_nanoscale_2014}.
In these studies, it seems that the efficiency of the heat engine
could be higher than the Carnot bound $\eta_{\text{\textsc{c}}}=1-T_{c}/T_{h}$.
However, since the baths are not canonical thermal ones, the parameter
$T$ can no longer be regarded as the well defined temperature. As
we have emphasized, such kind of systems are indeed not within the
applicable scope of the standard thermodynamics, therefore they do
not need to obey the second law inequalities that are based on canonical
thermal baths \citep{gardas_thermodynamic_2015}.

In this non-thermal bath case, the thermal entropy $dS_{\mathrm{e}}=\text{\text{\text{\dj}}}Q/T$
does not apply, but the information entropy is still well defined.
Now we study the system-bath mutual information when the baths are
non-thermal states. We consider an example of a single mode boson
($\hat{H}_{\text{\textsc{s}}}=\Omega\hat{a}^{\dagger}\hat{a}$) which
is linearly coupled with multiple squeezed thermal baths ($\hat{H}_{\text{\textsc{b}}}=\sum_{\alpha}\hat{H}_{\text{\textsc{b}},\alpha}$
and $\hat{H}_{\text{\textsc{b}},\alpha}=\sum_{k}\omega_{\alpha k}\,\hat{b}_{\alpha k}^{\dagger}\hat{b}_{\alpha k}$),
and they interact through $\hat{V}_{\text{\textsc{sb}}}=\sum_{\alpha}g_{\alpha k}\hat{a}^{\dagger}\hat{b}_{\alpha k}+g_{\alpha k}^{*}\hat{a}\hat{b}_{\alpha k}^{\dagger}$.
The initial states of the baths are squeezed thermal ones,
\begin{gather}
\hat{\rho}_{\text{\textsc{b}},\alpha}(0)=\frac{1}{Z_{\alpha}}\exp\big[-\frac{1}{T_{\alpha}}\,\hat{{\cal S}}_{\alpha}\hat{H}_{\text{\textsc{b}},\alpha}\hat{{\cal S}}_{\alpha}^{\dagger}\big],\label{eq:bath-Sq}\\
\hat{{\cal S}}_{\alpha}:=\prod_{k}\exp[\frac{1}{2}\lambda_{\alpha k}^{*}\hat{b}_{\alpha k}^{2}-\mathbf{h.c.}],\quad\lambda_{\alpha k}=r_{\alpha k}e^{-i\theta_{\alpha k}},\nonumber 
\end{gather}
 where $\hat{{\cal S}}_{\alpha}$ is the squeezing operator for bath-$\alpha$.
Below we will use the master equation to calculate the Spohn formula
(\ref{eq:Spohn}), and compare it with the result by directly calculating
the bath entropy change. We will see, in this non-thermal case, the
Spohn formula (\ref{eq:Spohn}) is still equal to the increasing rate
of the correlation between the system and the squeezed baths.

\textbf{\vspace{0.2em}\\ \noindent  Master equation: }We first look
at the dynamics of the open system alone. The total \noun{s+b }system
follows the von Neumann equation $\partial_{t}\hat{\rho}_{\text{\textsc{sb}}}(t)=i[\hat{\rho}_{\text{\textsc{sb}}}(t),\,\hat{{\cal H}}_{\text{\textsc{s+b}}}]$.
Based on it, after the Born-Markovian approximation \citep{breuer_theory_2002,walls_quantum_2008},
we can derive a master equation $\dot{\hat{\rho}}_{\text{\textsc{s}}}=\sum_{\alpha}{\cal L}_{\alpha}[\hat{\rho}_{\text{\textsc{s}}}]$
for the open system $\hat{\rho}_{\text{\textsc{s}}}(t)$ (interaction
picture), where (see the detailed derivation in Ref.\,\citep{li_production_2017})
\begin{align}
{\cal L}_{\alpha}[\hat{\rho}_{\text{\textsc{s}}}]=\gamma_{\alpha} & \Big[\mathfrak{n}_{\alpha}\big(\hat{a}^{\dagger}\hat{\rho}_{\text{\textsc{s}}}\hat{a}-\frac{1}{2}\{\hat{a}\hat{a}^{\dagger},\hat{\rho}_{\text{\textsc{s}}}\}\big)+(\mathfrak{n}_{\alpha}+1)\big(\hat{a}\hat{\rho}_{\text{\textsc{s}}}\hat{a}^{\dagger}-\frac{1}{2}\{\hat{a}^{\dagger}\hat{a},\hat{\rho}_{\text{\textsc{s}}}\}\big)\nonumber \\
 & -\mathfrak{u}_{\alpha}\big(\hat{a}^{\dagger}\hat{\rho}_{\text{\textsc{s}}}\hat{a}^{\dagger}-\frac{1}{2}\{(\hat{a}^{\dagger})^{2},\hat{\rho}_{\text{\textsc{s}}}\}\big)-\mathfrak{u}_{\alpha}^{*}\big(\hat{a}\hat{\rho}_{\text{\textsc{s}}}\hat{a}-\frac{1}{2}\{\hat{a}^{2},\hat{\rho}_{\text{\textsc{s}}}\}\big)\Big].
\end{align}
Here $\mathfrak{n}_{\alpha}:=(\overline{\mathrm{n}}_{\alpha,\Omega}+\frac{1}{2})\cosh2r_{\alpha\Omega}-\frac{1}{2}$,
$\mathfrak{u}_{\alpha}:=e^{i\theta_{\alpha\Omega}}(\overline{\mathrm{n}}_{\alpha,\Omega}+\frac{1}{2})\sinh2r_{\alpha\Omega}$,
and $\overline{\mathrm{n}}_{\alpha,\Omega}:=[\exp(\Omega/T_{\alpha})-1]^{-1}$
is the Planck function. The decay factor $\gamma_{\alpha}:=J_{\alpha}(\Omega)=K_{\alpha}(\Omega)$
is defined from the coupling spectrums $J_{\alpha}(\omega):=2\pi\sum_{k}|g_{\alpha k}|^{2}\delta(\omega-\omega_{\alpha k})$
and $K_{\alpha}(\omega):=2\pi\sum_{k}g_{\alpha k}^{2}\delta(\omega-\omega_{\alpha k})$.
And we have omitted the phase of $g_{\alpha k}$, thus $K_{\alpha}(\omega)=K_{\alpha}^{*}(\omega)=J_{\alpha}(\omega)$.
From this master equation, we obtain
\begin{gather}
\frac{d}{dt}\langle\tilde{a}(t)\rangle=-\sum_{\alpha}\frac{1}{2}\gamma_{\alpha}\langle\tilde{a}\rangle,\quad\frac{d}{dt}\langle\tilde{a}^{2}\rangle=-\sum_{\alpha}\gamma_{\alpha}[\langle\tilde{a}^{2}\rangle-\mathfrak{u}_{\alpha}],\nonumber \\
\frac{d}{dt}\langle\tilde{a}^{\dagger}\tilde{a}\rangle=-\sum_{\alpha}\gamma_{\alpha}[\langle\tilde{a}^{\dagger}\tilde{a}\rangle-\mathfrak{n}_{\alpha}].\label{eq:dynamics}
\end{gather}
Here $\langle\tilde{o}(t)\rangle:=\mathrm{tr}[\hat{\rho}_{\text{\textsc{s}}}\hat{o}(t)]$
gives variables in the rotating frame\footnote{Here $\hat{\rho}_{\text{\textsc{s}}}$
is in the interaction picture, and $\hat{o}$ is in the Schr{\"o}dinger
picture, thus we have { $\langle\hat{a}(t)\rangle=\langle\tilde{a}(t)\rangle e^{-i\Omega t}$},
where {$\langle\hat{o}(t)\rangle$} is the observable expectation
independent of pictures, and {$\langle\tilde{o}(t)\rangle$} is
the value in the rotating frame.}.

For this master equation, the partial steady state $\hat{\rho}_{\mathrm{ss}}^{(\alpha)}$
associated with bath-$\alpha$, which satisfies ${\cal L}_{\alpha}[\hat{\rho}_{\mathrm{ss}}^{(\alpha)}]=0$,
is a squeezed thermal one,
\begin{gather}
\hat{\rho}_{\mathrm{ss}}^{(\alpha)}=\frac{1}{\text{\textsc{z}}_{\alpha}}\exp[-\frac{\Omega}{T_{\alpha}}\cdot\hat{\mathsf{S}}_{\alpha}\hat{a}^{\dagger}\hat{a}\hat{\mathsf{S}}_{\alpha}^{\dagger}],\\
\hat{\mathsf{S}}_{\alpha}:=\exp[-(\frac{1}{2}\zeta_{\alpha}^{*}\hat{a}^{2}-\mathbf{h.c.})],\quad\zeta_{\alpha}=\lambda_{\alpha k}\big|_{\omega_{k}=\Omega}:=r_{\alpha}e^{i\theta_{\alpha}},\nonumber 
\end{gather}
 where $\hat{\mathsf{S}}_{\alpha}$ is the squeezing operator. Now
we can put this result into the Spohn formula (\ref{eq:Spohn}), then
the term $\chi_{\alpha}=\mathrm{tr}\big[{\cal L}_{\alpha}[\hat{\rho}]\cdot\ln\hat{\rho}_{\mathrm{ss}}^{(\alpha)}\big]$
gives 
\begin{equation}
\chi_{\alpha}=\frac{\Omega}{T_{\alpha}}\cdot\gamma_{\alpha}\Big(\cosh2r_{\alpha}\cdot[\langle\tilde{a}^{\dagger}\tilde{a}\rangle-\mathfrak{n}_{\alpha}]-\frac{1}{2}\sinh2r_{\alpha}[e^{-i\theta_{\alpha}}(\langle\tilde{a}^{2}(t)\rangle-\mathfrak{u}_{\alpha})+\mathbf{h.c.}]\Big).\label{eq:Xa}
\end{equation}

When there is no squeezing ($r_{\alpha}=0$), this equation exactly
returns to the thermal bath result $\chi_{\alpha}=-\frac{1}{T_{\alpha}}\frac{d}{dt}[\Omega\langle\tilde{a}^{\dagger}\tilde{a}\rangle]=-\dot{Q}_{\alpha}/T_{\alpha}$
{[}see Eq.\,(\ref{eq:dynamics}){]}, which is the exchange of the
thermal entropy. However, due to the quantum squeezing in the bath,
clearly this $\chi_{\alpha}$ term is no longer the thermal entropy,
and now it looks too complicated to tell its physical meaning. Below,
we are going to show that here this $\chi_{\alpha}$ term is just
the informational entropy changing of bath-$\alpha$.

\textbf{\vspace{0.2em}\\ \noindent } \textbf{Bath entropy dynamics:
}Now we calculate the entropy change $\dot{S}_{\text{\textsc{b}},\alpha}$
of bath-$\alpha$ directly by adopting the similar approximation as
Eq.\,(\ref{eq:S_Ba}), and that gives
\begin{align}
\dot{S}_{\text{\textsc{b}},\alpha} & \simeq-\mathrm{tr}\big[\dot{\hat{\rho}}_{\text{\textsc{b}},\alpha}(t)\ln\Big(\frac{1}{Z_{\alpha}}\exp\big[-\frac{1}{T_{\alpha}}\,\hat{{\cal S}}_{\alpha}\hat{H}_{\text{\textsc{b}},\alpha}\hat{{\cal S}}_{\alpha}^{\dagger}\big]\Big)\big]\nonumber \\
 & =\sum_{k}\frac{\omega_{\alpha k}}{T_{\alpha}}\Big(\cosh2r_{\alpha k}\cdot\frac{d}{dt}\langle\tilde{b}_{\alpha k}^{\dagger}(t)\tilde{b}_{\alpha k}(t)\rangle+\frac{1}{2}\sinh2r_{\alpha k}[e^{-i\theta_{\alpha k}}\cdot\frac{d}{dt}\langle\tilde{b}_{\alpha k}^{2}(t)\rangle+\mathbf{h.c.}]\Big).\label{eq:S_Ba_sq}
\end{align}

Unlike the thermal baths case in Eq.\,(\ref{eq:S_Ba}), here it is
not easy to see how the bath entropy dynamics $\dot{S}_{\text{\textsc{b}},\alpha}$
is related the system dynamics. But notice that $\dot{S}_{\text{\textsc{b}},\alpha}$\textbf{
}is simply determined by the time derivative of the bath operator
expectations like $\langle\tilde{b}_{\alpha k}^{\dagger}(t)\tilde{b}_{\alpha k}(t)\rangle$
and $\langle\tilde{b}_{\alpha k}^{2}(t)\rangle$, which can be further
calculated by Heisenberg equations. After certain Markovian approximation,
in the weak coupling limit ($\gamma_{\alpha}\ll\Omega$), we can prove
the following relation (see the detailed proof in Ref.\,\citep{li_production_2017}),
\begin{align}
\sum_{k}\mathfrak{f}_{k}\cdot\frac{d}{dt}\langle\tilde{b}_{\alpha k}^{\dagger}\tilde{b}_{\alpha k}\rangle & \simeq\mathfrak{f}(\omega_{k}\rightarrow\Omega)\cdot\gamma_{\alpha}[\langle\tilde{a}^{\dagger}\tilde{a}\rangle-\mathfrak{n}_{\alpha}]=-\mathrm{tr}\Big\{\mathfrak{f}(\Omega)\tilde{a}^{\dagger}\tilde{a}\cdot{\cal L}_{\alpha}[\rho]\Big\},\label{eq:F-k}\\
\sum_{k}\mathfrak{h}_{k}\cdot\frac{d}{dt}\langle\tilde{b}_{\alpha k}^{2}\rangle & \simeq-\mathfrak{h}(\omega_{k}\rightarrow\Omega)\cdot\gamma_{\alpha}[\langle\tilde{a}^{2}\rangle-\mathfrak{u}_{\alpha}]=\mathrm{tr}\Big\{\mathfrak{h}(\Omega)\tilde{a}^{2}\cdot{\cal L}_{\alpha}[\rho]\Big\},\nonumber 
\end{align}
where $\mathfrak{f}_{k}$ and $\mathfrak{h}_{k}$ are the summation
weights associated with the bath mode $\hat{b}_{\alpha k}$. 

These two relations well connects the dynamics of bath-$\alpha$ (left
sides) with that of the open system (right sides). For example, let
$\mathfrak{f}_{k}=\omega_{\alpha k}$, then the above relation becomes
$\frac{d}{dt}[\sum_{k}\omega_{\alpha k}\langle\tilde{b}_{\alpha k}^{\dagger}\tilde{b}_{\alpha k}\rangle]\simeq-\mathrm{tr}\big(\Omega\tilde{a}^{\dagger}\tilde{a}\cdot{\cal L}_{\alpha}[\rho]\big)$,
which is just the heat emission-absorption relation $\frac{d}{dt}\langle\hat{H}_{\text{\textsc{b}},\alpha}\rangle=-\dot{Q}_{\alpha}$,
and we have utilized it in the discussion below Eq.\,(\ref{eq:S_Ba}).
To calculate the above entropy change Eq.\,(\ref{eq:S_Ba_sq}) for
the squeezed thermal bath, let $\mathfrak{f}_{k}=\frac{1}{T_{\alpha}}\omega_{\alpha k}\cosh2r_{\alpha k}$,
$\mathfrak{h}_{k}=\frac{1}{2T_{\alpha}}\omega_{\alpha k}e^{-i\theta_{\alpha k}}\sinh2r_{\alpha k}$,
then we obtain 
\begin{equation}
\dot{S}_{\text{\textsc{b}},\alpha}=\frac{\Omega}{T_{\alpha}}\cdot\gamma_{\alpha}\Big(\cosh2r_{\alpha}\cdot[\langle\tilde{a}^{\dagger}\tilde{a}\rangle-\mathfrak{n}_{\alpha}]-\frac{1}{2}\sinh2r_{\alpha}[e^{-i\theta_{\alpha}}(\langle\tilde{a}^{2}(t)\rangle-\mathfrak{u}_{\alpha})+\mathbf{h.c.}]\Big).\label{eq:sigma_B}
\end{equation}

This result exactly equals to the term $\chi_{\alpha}=\mathrm{tr}\big[{\cal L}_{\alpha}[\rho]\ln\rho_{\mathrm{ss}}^{(\alpha)}\big]$
in the above Spohn formula {[}see Eq.\,(\ref{eq:Xa}){]}. Therefore,
in this non-thermal bath case, the changing rate of the system-bath
mutual information is just equal to the Spohn formula (\ref{eq:Spohn}),
\begin{equation}
\frac{d}{dt}{\cal I}_{\text{\textsc{sb}}}=\dot{S}_{\text{\textsc{s}}}+\sum_{\alpha}\dot{S}_{\text{\textsc{b}},\alpha}=R_{\mathrm{Sp}}\ge0,
\end{equation}
 thus its positivity is still guaranteed \citep{li_production_2017}. 

That means, although the non-thermal baths are beyond the applicable
scope of the standard thermodynamics, namely, $dS_{\mathrm{i}}=dS-\sum_{\alpha}\text{\text{\text{\dj}}}Q_{\alpha}/T_{\alpha}\ge0$
does not apply, the system-bath correlation ${\cal I}_{\text{\textsc{sb}}}$
still keeps increasing monotonically like in the thermal bath case
(Sec.\,\ref{subsec:S-B-correlation}). Therefore, this system-bath
correlation production may be a generalization for the irreversible
entropy production which also applies for the non-thermal cases. Tracing
back to the original consideration of the irreversible entropy change
{[}Eq.\,(\ref{eq:dSi-0}){]}, it turns out the term $-dS_{\mathrm{e}}$
can also be regarded the informational entropy change of the bath,
and it gives the relation $dS_{\mathrm{e}}=\text{\text{\text{\dj}}}Q/T$
in the special case of canonical thermal bath.

\subsection{Discussions}

Historically the Spohn formula was first introduced by considering
the distance between the system state $\hat{\rho}(t)$ and its final
steady state $\hat{\rho}_{\mathrm{ss}}$, which is measured by their
relative entropy $S[\hat{\rho}(t)\parallel\hat{\rho}_{\mathrm{ss}}]:=-\mathrm{tr}\big[\hat{\rho}(t)\cdot(\ln\hat{\rho}(t)-\ln\hat{\rho}_{\mathrm{ss}})\big]$
(Spohn \citep{spohn_entropy_1978}). When $t\rightarrow\infty$, $\hat{\rho}(t)\rightarrow\hat{\rho}_{\mathrm{ss}}$,
and this distance decreases to zero. Thus, for a Markovian master
equation $\partial_{t}\hat{\rho}={\cal L}[\hat{\rho}]$, the EPr is
defined from the time derivative of this distance, i.e.,
\begin{equation}
\sigma:=-\frac{d}{dt}S[\hat{\rho}(t)\parallel\hat{\rho}_{\mathrm{ss}}]=\mathrm{tr}\big[(\ln\hat{\rho}_{\mathrm{ss}}-\ln\hat{\rho}){\cal L}[\hat{\rho}]\big].
\end{equation}
Therefore, this EPr-$\sigma$ serves as a Lyapunov index for the master
equation. It was proved that $\sigma\ge0$, and $\sigma=0$ when $t\rightarrow\infty$.
When there is only one thermal bath, this EPr-$\sigma$ returns to
the thermodynamics result, $\sigma=\dot{S}-\dot{Q}/T$.

However, when the open system contacts with multiple heat baths as
described by the master equation (\ref{eq:QME}), denoting ${\cal L}[\hat{\rho}]=\sum_{\alpha}{\cal L}_{\alpha}[\hat{\rho}]$,
the above EPr-$\sigma$ still goes to zero when $t\rightarrow\infty$.
Thus it does not tell the difference between achieving the equilibrium
state or the stationary non-equilibrium state {[}see the example of
Eq.\,(\ref{eq:T1T2}){]}. Later (Spohn, Lebowitz \citep{spohn_irreversible_1978}),
this EPr-$\sigma$ was generalized to be the form of Eqs.\,(\ref{eq:R-EPr},
\ref{eq:Spohn}), and its positivity can be proved by the similar
procedure given in the previous study (Spohn \citep{spohn_entropy_1978}).

In the standard thermodynamics, the second law statement $dS_{\mathrm{i}}\ge0$
requires a \emph{monotonic} increase of the irreversible entropy,
not only comparing with the initial state. Notice that in the proof
for the positivity of the EPr, the Markovianity is necessary for both
classical and quantum cases. If the master equation of the open system
is non-Markovian, it is possible that there exist certain periods
where $R_{\text{\textsc{ep}}}<0$, which means the decrease of the
irreversible entropy (or the system-bath correlation). When comparing
EPr with the standard thermodynamics, a coarse-grained time scale
is more proper (which usually means the Markovian process), thus $R_{\text{\textsc{ep}}}<0$
may be acceptable if it appears only in short time scales.

In the above discussions, clearly the most important part is how to
calculate the bath entropy dynamics directly. This is usually quite
difficult since the bath contains infinite DoF. The above calculation
can be done mainly thanks to the approximation $\dot{S}_{\text{\textsc{b}}}\simeq-\mathrm{tr}[\dot{\hat{\rho}}_{\text{\textsc{b}}}(t)\ln\hat{\rho}_{\text{\textsc{b}}}(0)]$.
The results derived thereafter are consistent with the previous conclusions
in thermodynamics, but still we need more examination about the validity
of this approximation.

There are few models of open system that are exactly solvable for
this examination. In Ref.\,\citep{you_entropy_2018}, the bath entropy
dynamics was calculated when a two-level-system (TLS) is dispersively
coupled with a squeezed thermal bath. In this problem, the density
matrix evolution of each bath mode can be exactly solved. The state
of each bath mode is the probabilistic summation of two displaced
Gaussian states $\hat{\varrho}_{k}(t)=p_{+}\hat{\varrho}_{k}^{+}(t)+p_{-}\hat{\varrho}_{k}^{-}(t)$,
which keep separating and recombining periodically in the phase space.
Thus the exact entropy dynamics can be calculated and compared with
the result based on the above approximation. 

It turns out the above approximation fits the exact result quite well
in the high temperature area; in the low temperature area, the approximated
result diverges to infinity when $T\rightarrow0$, but the exact result
remains finite. This is because in the high temperature area, this
$\hat{\varrho}_{k}(t)$ can be better approximated as a single Gaussian
state when the separation of $\hat{\varrho}_{k}^{\pm}(t)$ is quite
small; while in the low temperature area, the uncertainty of $\hat{\varrho}_{k}(t)$
mainly comes from the probabilities $p_{\pm}$ but not the entropy
in the Gaussian states $\hat{\varrho}_{k}^{\pm}$. Namely, in the
low temperature area, the influence from the system to the bath is
bigger, especially for nonlinear systems like the TLS. If the bath
states cannot be well treated as Gaussian ones, the above approximation
is questionable, and how to calculate the bath entropy in this case
remains an open problem.

\section{The entropy in the ideal gas diffusion }

In the above discussions about the correlation production between
the open system and its environment, we utilized an important condition,
i.e., the whole \noun{s+b }system is an isolated system, thus its
entropy does not change with time. In the quantum case, the whole
\noun{s+b }system follows the von Neumann equation $\partial_{t}\hat{\rho}_{\text{\textsc{sb}}}=i[\hat{\rho}_{\text{\textsc{sb}}},\,\hat{{\cal H}}_{\text{\textsc{s+b}}}]$,
thus the von Neumann entropy $S_{\text{\textsc{v}}}[\hat{\rho}_{\text{\textsc{sb}}}]=-\mathrm{tr}[\hat{\rho}_{\text{\textsc{sb}}}\ln\hat{\rho}_{\text{\textsc{sb}}}]$
does not change during the unitary evolution. Likewise, in the classical
case, the whole system follows the Liouville equation $\partial_{t}\rho_{\text{\textsc{sb}}}=-\{\rho_{\text{\textsc{sb}}},\,{\cal H}_{\text{\textsc{s+b}}}\}$,
thus the Gibbs entropy $S_{\text{\textsc{g}}}[\rho]=-\sum_{n}p_{n}\ln p_{n}$
keeps a constant.

But still this is quite counter-intuitive comparing with our intuition
of the macroscopic irreversibility. For example, considering the diffusion
process of an ideal gas as we mentioned in the very beginning (Fig.\,\ref{fig-expansion}),
although there are no particle interactions and the dynamics of the
whole system is well predictable, still we could see the diffusion
proceeds irreversibly, and would finally occupy the whole volume uniformly.

In this section, we will show this puzzle also can be understood in
the sense of correlation production. In open systems, we have seen
it is the system-bath correlation that increases, while the total
\noun{s+b }entropy does not change. In an isolated system, there is
no partition for ``system'' and ``bath'', but we will see it is
the correlation between different DoF, e.g., position-momentum, and
particle-particle, that increases monotonically, while the total entropy
does not change \citep{hobson_irreversibility_1966,hobson_concepts_1971}.

\subsection{Liouville dynamics of the ideal gas diffusion}

Here we make a full calculation on the phase-space evolution of the
above ideal gas diffusion process in classical physics, so as to examine
the dynamical behavior of the microstate, as well as its entropy. 

Since there is no interaction between particles, the dynamics of the
$3N$ DoF are independent from each other, the total $N$-particle
microstate PDF can be written as a product form, i.e., $\rho(\vec{P},\vec{Q},t)=\prod_{i,\sigma}\varrho(p_{i}^{\sigma},q_{i}^{\sigma},t)$
($\sigma=x,y,z$) \footnote{Assuming there is no initial correlation
between different DoF.}, thus this problem can be reduced to study
the PDF of a single DoF $\varrho(p,x,t)$. Correspondingly, the Liouville
equation is 
\begin{equation}
\partial_{t}\varrho=-\{\varrho,\,H\}=-\frac{\partial\varrho}{\partial x}\frac{\partial H}{\partial p}+\frac{\partial\varrho}{\partial p}\frac{\partial H}{\partial x}=-\frac{p}{m}\partial_{x}\varrho,\label{eq:Liouville}
\end{equation}
where $H=p^{2}/2m$ is the single DoF Hamiltonian.

This equation is exactly solvable, and the general solution is $\Phi(p,\,x-\frac{p}{m}t)$.
The detailed form of the function $\Phi(\cdots,\,\cdots)$ should
be further determined by the initial and boundary conditions. We assume
initially the system starts from an equilibrium state confined in
the area $x\in[a,b]$, namely 
\begin{equation}
\varrho(p,x,0)=\Lambda(p)\times\Pi(x).\label{eq:initial}
\end{equation}
Here $\Lambda(p)=\frac{1}{Z}\exp[-p^{2}/2\bar{p}_{T}^{2}]$ is the
MB distribution, with $\bar{p}_{T}^{2}/2m=\frac{1}{2}k_{\text{\textsc{b}}}T$
as the average kinetic energy, and $Z=\sqrt{2\pi}\,\bar{p}_{T}$ is
a normalization factor. $\Pi$(x) is the initial spatial distribution
{[}Fig.\,\ref{fig-phasePDF}(a){]}
\begin{equation}
\Pi(x)=\begin{cases}
\frac{1}{b-a}, & a\le x\le b,\\
0, & \text{elsewhere}.
\end{cases}
\end{equation}
Such a product form of $\varrho(p,x,0)$ indicates the spatial and
momentum distributions have no correlations in priori.

For the diffusion in free space $x\in(-\infty,\infty)$, the time-dependent
solution is 
\begin{equation}
\varrho_{\text{\textsc{f}}}(p,x,t)=\Lambda(p)\Pi(x-\frac{p}{m}t),\label{eq:free}
\end{equation}
which satisfies both the Liouville equation (\ref{eq:Liouville})
and the initial condition (\ref{eq:initial}). 

\begin{figure}
\begin{centering}
\includegraphics[width=0.5\textwidth]{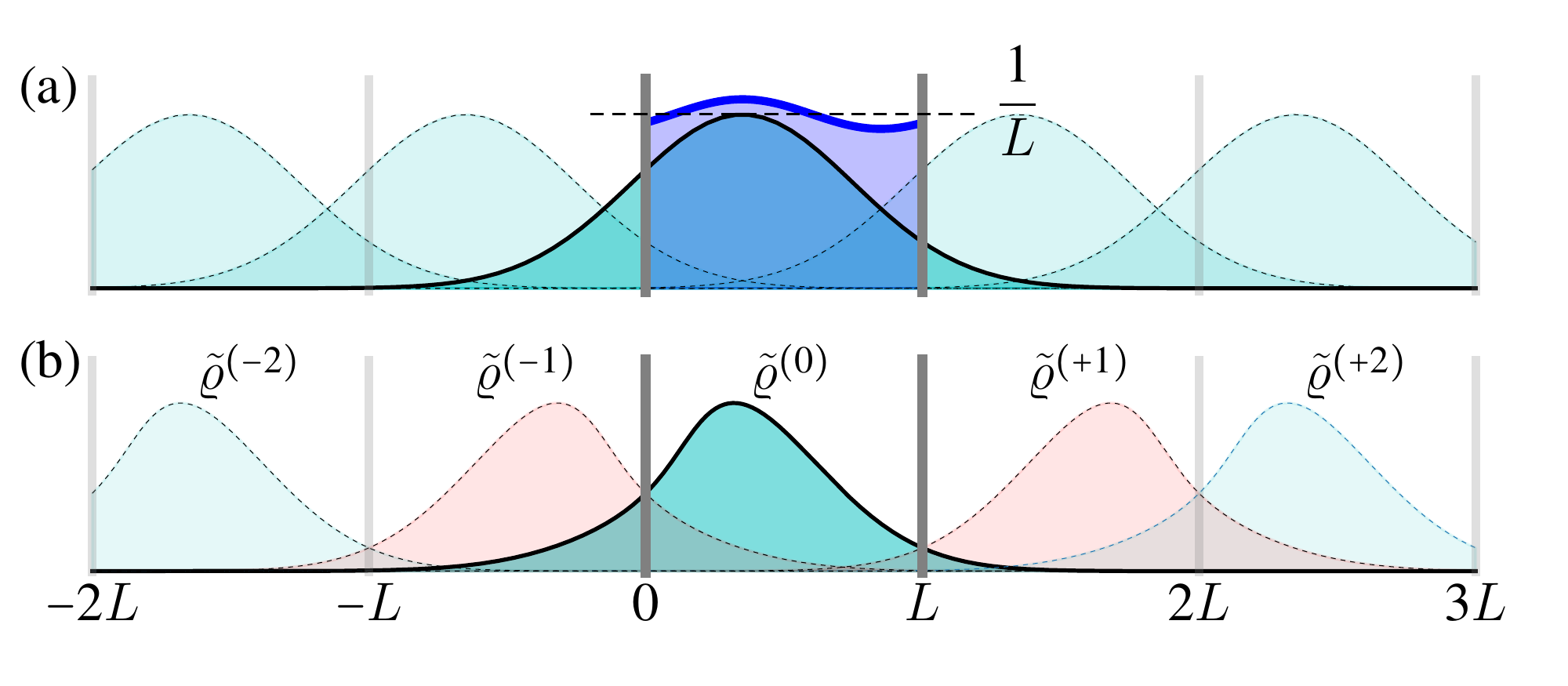}
\par\end{centering}
\caption{Demonstration for how the solutions are constructed for (a) periodic
and (b) reflecting boundary conditions. The free space is cut into
intervals of length $L$, and each contributes an ``image'' source.
The solution is all their summation in $x\in[0,L]$ {[}blue line in
(a){]}.}

\label{fig-interval}
\end{figure}

For a confined area $x\in[0,L]$ with periodic boundary condition
$\varrho(p,0,t)=\varrho(p,L,t)$, the solution can be constructed
with the help of the above free space one, i.e.,
\begin{equation}
\varrho(p,x,t)=\sum_{n=-\infty}^{\infty}\varrho_{\text{\textsc{f}}}(p,x+nL,t),\quad0\le x\le L.\label{eq:periodic}
\end{equation}
Here $\varrho_{\text{\textsc{f}}}(p,x+nL,t)$ can be regarded as the
periodic ``image'' solution in the interval $[nL,nL+L]$ {[}Fig.\,\ref{fig-interval}(a){]}
\citep{wang_magnetic_2018}. Clearly, Eq.\,(\ref{eq:periodic}) satisfies
the periodic boundary condition, as well as the initial condition
(\ref{eq:initial}), and it is simple to verify each summation term
satisfies the above Liouville equation (\ref{eq:Liouville}), thus
Eq.\,(\ref{eq:periodic}) describes the full microstate PDF evolution
in the confined area $x\in[0,L]$ with periodic boundary condition.

From these exact solutions (\ref{eq:free}, \ref{eq:periodic}), it
is clear to see the microstate PDF $\varrho(p,x,t)$ can no longer
hold the separable form like $f_{\mathtt{x}}(x,t)\times f_{\mathtt{p}}(p,t)$
once the diffusion starts, thus indeed it is not evolving towards
any equilibrium state, since an equilibrium state must have a separable
form similar like the initial condition (\ref{eq:initial}) {[}see
also Fig.\,\ref{fig-phasePDF}(a){]}.

In Fig.\,\ref{fig-phasePDF} we show the microstate PDF $\varrho(p,x,t)$
at different times. As the time increases, the ``stripe'' in Fig.\,\ref{fig-phasePDF}(a)
becomes more and more inclined; once exceeding the boundary, it winds
back from the other side due to the periodic boundary condition and
generates a new ``stripe'' {[}Fig.\,\ref{fig-phasePDF}(c){]}.
After very long time, more and more stripes appear, much denser and
thinner, but they would never occupy the whole phase space continuously
{[}Fig.\,\ref{fig-phasePDF}(d, e){]}. 

Fig.\,\ref{fig-phasePDF}(e) shows the conditional PDF of the momentum
when the position is fixed at $x=L/2$ {[}the vertical dashed line
in Fig.\,\ref{fig-phasePDF}(d){]}. In the limit $t\rightarrow\infty$,
it becomes an exotic function \emph{discontinuous everywhere}, but
not the MB distribution. All these features indicate that, during
this diffusion process of the isolated ideal gas, indeed the ensemble
is not evolving towards the new equilibrium state as expected in the
macroscopic intuition. Even after long time relaxation, the microstate
PDF $\varrho(p,x,t\rightarrow\infty)$ is not approaching the equilibrium
state.

\subsection{Spatial and momentum distributions }

\begin{figure}
\begin{centering}
\includegraphics[width=0.5\textwidth]{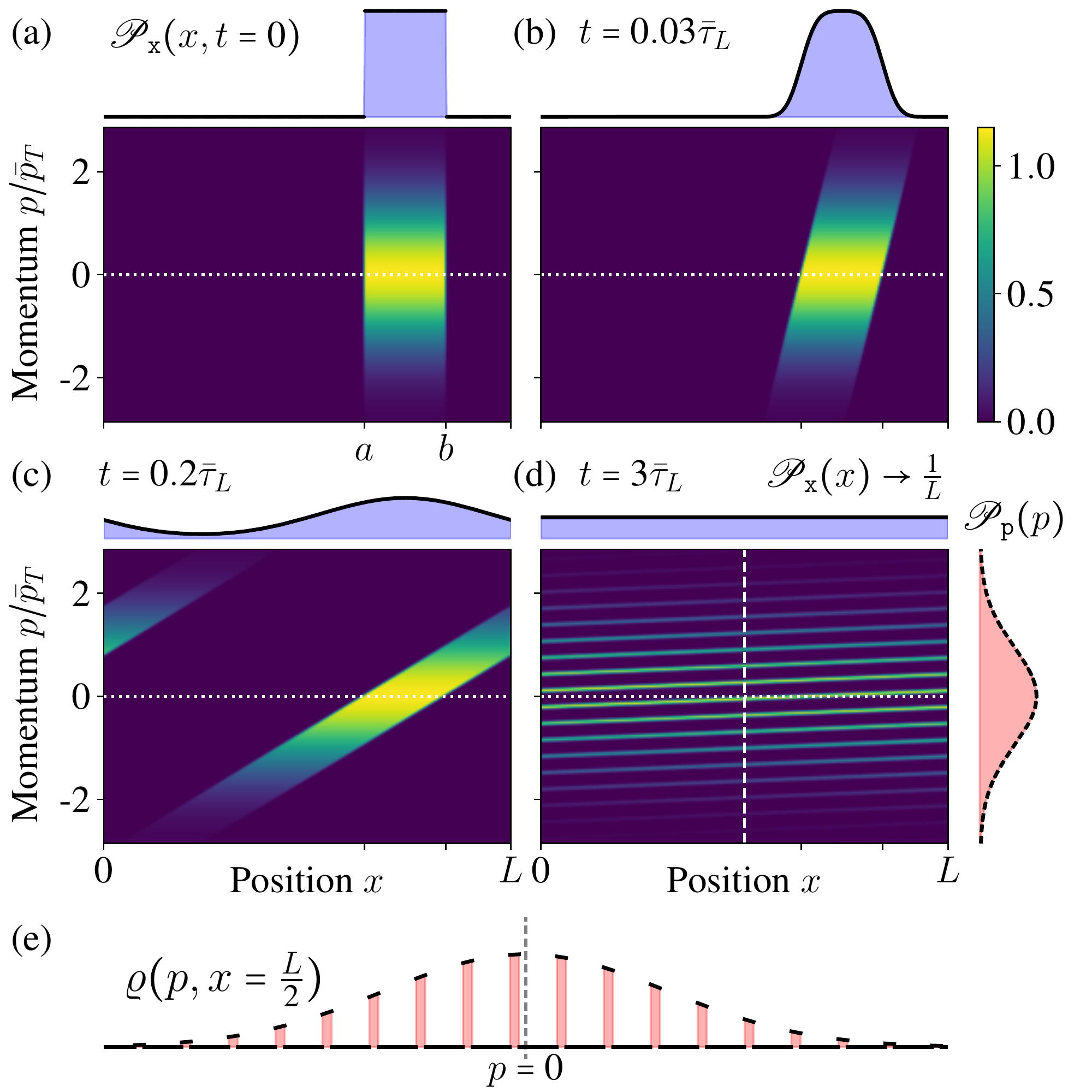}
\par\end{centering}
\caption{(Color online) (a-d) The distribution $\varrho(p,x,t)$ in phase space
at different times ($\bar{\tau}_{L}:=mL/\bar{p}_{T}$ as the time
unit). As the time increases, $\mathscr{P}_{\mathtt{p}}(p)$ does
not change, but $\mathscr{P}_{\mathtt{x}}(x,t)$ approaches the new
uniform distribution in $x\in[0,L]$. (e) The conditional distribution
at a fixed position $\varrho(p,x=L/2)$ {[}vertical dashed line in
(d){]}.}

\label{fig-phasePDF}
\end{figure}

Even after long time relaxation, the ideal gas would not achieve the
new equilibrium state. This result looks counter-intuitive, since
clearly we can see the particles spread all over the box uniformly
after long enough time relaxation. However, we must notice that the
fact ``spreading all over the box uniformly'' is implicitly focused
on the position distribution $\mathscr{P}_{\mathtt{x}}(x,t)$ alone,
but not the whole ensemble state $\varrho(p,x,t)$. As a marginal
distribution of $\varrho(p,x,t)$, the spatial distribution $\mathscr{P}_{\mathtt{x}}(x,t)\rightarrow1/L$
does approach the new uniform one as its steady state {[}Fig.\,\ref{fig-phasePDF}(d){]},
and now we show indeed this is true for any initial state of $\Pi(x)$
\citep{hobson_irreversibility_1966,hobson_concepts_1971}.

We first consider the initial spatial distribution is a $\delta$-function
concentrated at $x_{0}$, $\Pi(x)=\delta(x-x_{0})$. Since we have
obtained the analytical results (\ref{eq:free}, \ref{eq:periodic})
for the ensemble evolution $\varrho(p,x,t)$, the spatial distribution
$\mathscr{P}_{\mathtt{x}}(x,t)$ emerges as its marginal distribution
by averaging over the momentum: 
\begin{align}
\mathscr{P}_{\mathtt{x}}(x,t) & =\int_{-\infty}^{\infty}dp\,\varrho(p,x,t)=\int_{-\infty}^{\infty}dp\,\sum_{n=-\infty}^{\infty}\frac{1}{Z}\exp[-\frac{p^{2}}{2\bar{p}_{T}^{2}}]\times\delta(x+nL-x_{0}-\frac{p}{m}t)\nonumber \\
 & =\sum_{n=-\infty}^{\infty}\frac{m}{Zt}\exp\big[-\frac{1}{2\bar{v}_{T}^{2}t^{2}}(x+nL-x_{0})^{2}\big],\label{eq:Px}
\end{align}
where $\bar{v}_{T}:=\bar{p}_{T}/m$. With the increase of time $t$,
these Gaussian terms becomes wider and lower {[}Fig.\,\ref{fig-interval}(a){]}.
Therefore, when $t\rightarrow\infty$, the spatial distribution $\mathscr{P}_{\mathtt{x}}(x,t)$
always approaches the uniform distribution in $x\in[0,L]$.

Any initial spatial distribution can be regarded as certain combination
of $\delta$-functions, i.e., $\Pi(x)=\int dx_{0}\,\Pi(x_{0})\delta(x-x_{0})$.
Therefore, for any initial $\Pi(x)$, the spatial distribution $\mathscr{P}_{\mathtt{x}}(x,t)$
always approaches the uniform one as its steady state. In this sense,
although the underlying Liouville dynamics obeys the time-reversal
symmetry, the ``irreversible'' diffusion appears into our sight.

On the other hand, $\mathscr{P}_{\mathtt{p}}(p,t)$ never changes
with time, and always maintains its initial distribution, which can
be proved by simply changing the integral variable:
\begin{align}
\mathscr{P}_{\mathtt{p}}(p,t) & =\int_{0}^{L}dx\,\sum_{n=-\infty}^{\infty}\Lambda(p)\Pi(x+nL-\frac{p}{m}t)\nonumber \\
 & =\int_{-\infty}^{\infty}dx\,\Lambda(p)\Pi(x-\frac{p}{m}t)=\Lambda(p).
\end{align}
This is all because of the periodic boundary condition, and the particles
always move freely. If the particles can be reflected back at the
boundaries, this momentum distribution would also change with time.

\subsection{Reflecting boundary condition}

Now we consider the reflecting boundary condition. In this case, when
the particles hit the boundaries at $x=0,L$, their positions do not
change, but their momentum should be suddenly changed from  $p$ to
$-p$. Correspondingly, the analytical result for the ensemble evolution
can be obtained by summing up the ``reflection images'' {[}Fig.\,\ref{fig-interval}(b){]},
i.e., 
\begin{gather}
\varrho(p,x,t)=\tilde{\varrho}^{(0)}(p,x,t)+\sum_{n=1}^{\infty}\tilde{\varrho}^{(-n)}(p,x,t)+\tilde{\varrho}^{(+n)}(p,x,t),\qquad\text{for }x\in[0,L],\nonumber \\
\tilde{\varrho}^{(-n)}(p,x,t):=\mathbf{R}_{0}\big[\tilde{\varrho}^{[+(n-1)]}(-p,x,t)\big],\qquad\tilde{\varrho}^{(+n)}(p,x,t):=\mathbf{R}_{L}\big[\tilde{\varrho}^{[-(n-1)]}(-p,x,t)\big].
\end{gather}
 Here $\tilde{\varrho}^{(0)}(p,x,t)=\varrho_{\text{\textsc{f}}}(p,x,t)$,
and $\mathbf{R}_{a}[f(p,x)]:=f(p,2a-x)$ means making a mirror reflection
to the function $f(p,x)$ along the axis $x=a$. Clearly, each summation
term $\tilde{\varrho}^{(-n)}$ can be regarded as shifted from $\varrho_{\text{\textsc{f}}}(p,x,t)$
or its mirror reflection, thus they all satisfy the differential relation
in the Liouville equation (\ref{eq:Liouville}).

\begin{figure}
\begin{centering}
\includegraphics[width=0.5\textwidth]{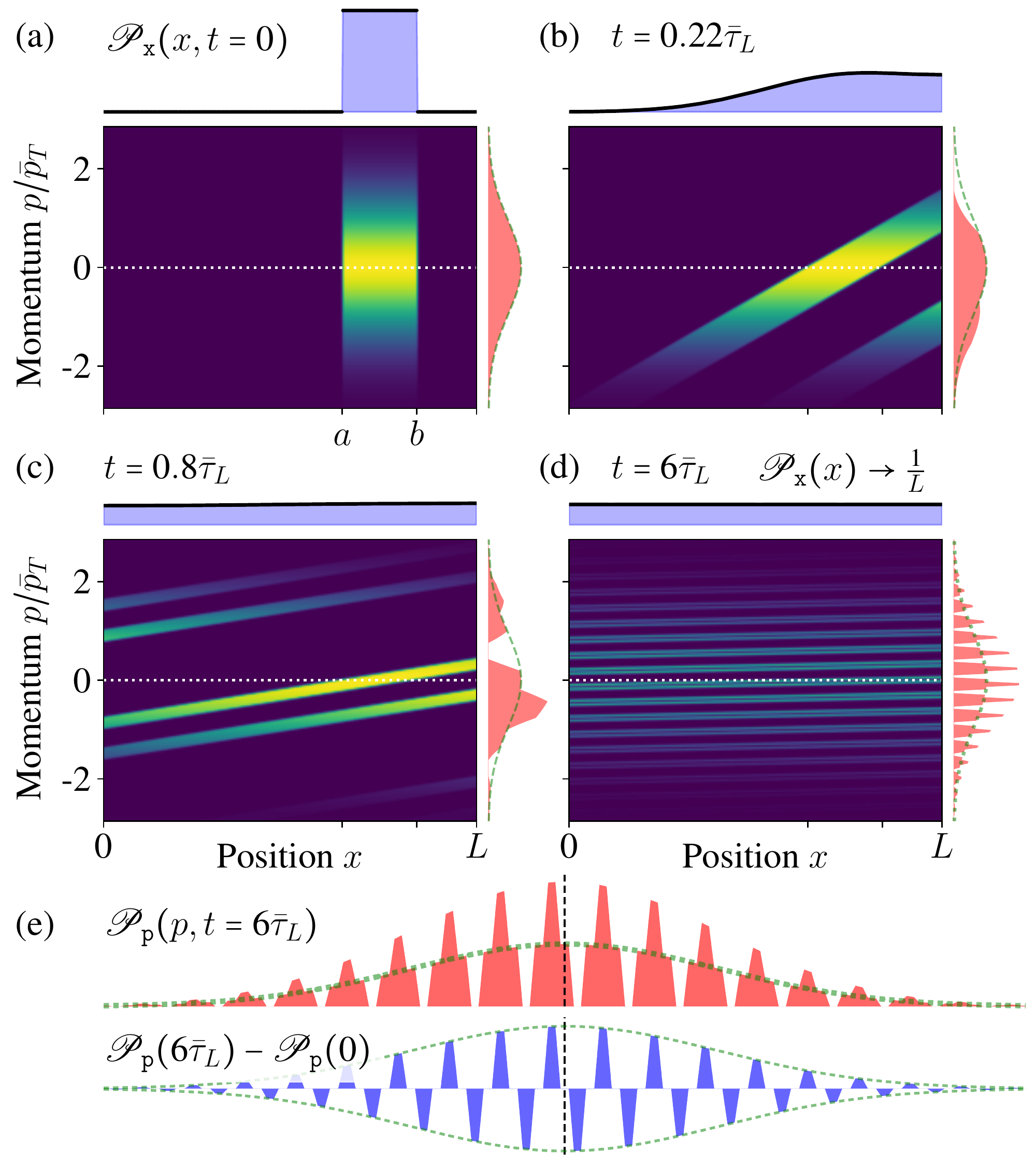}
\par\end{centering}
\caption{(Color online) (a-d) $\varrho(p,x,t)$ under reflecting boundary condition,
as well as its spatial and momentum distributions. (e) The momentum
distribution from (d), and its difference (lower blue) with the initial
MB one (green dashed lines).}

\label{fig-reflect}
\end{figure}
To verify the boundary condition, consider a diffusing distribution
in $x\in[0,L]$, initially described by $\tilde{\varrho}^{(0)}(p,x,t)$.
As the time increases, it diffuses wider and even exceeds the box
range $[0,L]$. The exceeded part should be reflected at the boundaries
$x=0,L$ as the next order $\tilde{\varrho}^{(\pm1)}$, and added
back to the total result $\varrho(p,x,t)$. This procedure should
be done iteratively, namely, when the term $\tilde{\varrho}^{(\pm n)}$
exceeds the boundaries, it generates the reflected term $\tilde{\varrho}^{[\mp(n+1)]}$
as the next summation order {[}Fig.\,\ref{fig-interval}(b){]}. 

Therefore, this result is quite similar to the above periodic case,
except reflection should be made to certain summation terms. Based
on the same reason as above, each summation term becomes more and
more flat during the diffusion, thus the spatial distribution $\mathscr{P}_{\mathtt{x}}(x,t)$
always approaches the new uniform one as its steady state \citep{hobson_concepts_1971}.

The ensemble evolution is shown in Fig.\,\ref{fig-reflect}(a-d),
which is quite similar with the above periodic case. Again, $\varrho(p,x,t)$
is indeed not evolving towards the equilibrium state. A significant
difference is the momentum distribution $\mathscr{P}_{\mathtt{p}}(p,t)$
now varies with time. This is because the collision at the boundaries
changes the momentum direction, thus $\langle p\rangle$ is no longer
conserved, although the kinetic energy $\langle p^{2}\rangle$ does
not change (as the momentum amplitude). 

Notice that the reflection ``moves'' the probability of momentum
$p$ to the area of $-p$, therefore, we see that some ``areas''
of $\mathscr{P}_{\mathtt{p}}(p,t)$ are ``cut'' off from the initial
MB distribution, and ``added'' to its mirror position along $p=0$
{[}especially Fig.\,\ref{fig-reflect}(c, d){]}. Thus $\mathscr{P}_{\mathtt{p}}(p,t)$
is different from the initial MB distribution. 

Since the reflection transfers the probability of $p$ to its mirror
position $-p$, the difference $\delta\mathscr{P}_{\mathtt{p}}(t):=\mathscr{P}_{\mathtt{p}}(t)-\mathscr{P}_{\mathtt{p}}(0)$
is always an odd function {[}lower blue in Fig.\,\ref{fig-reflect}(e){]}.
As a result, the even moments $\langle p^{2n}\rangle$ of $\mathscr{P}_{\mathtt{p}}(t)$
are the same with the MB distribution, but the odd ones $\langle p^{2n+1}\rangle$
are changed.

As the time increases, more and more ``stripes'' appear in $\mathscr{P}_{\mathtt{p}}(p,t)$,
much thinner and denser. As a result, when calculating the odd orders
$\langle p^{2n+1}\rangle$, the contributions from the nearest two
stripes in $\delta\mathscr{P}_{\mathtt{p}}(t)$ (who have similar
$p$ values), positive and negative, tends to cancel each other {[}lower
blue in Fig.\,\ref{fig-reflect}(e){]}. Therefore, in the limit $t\rightarrow\infty$,
the odd orders $\langle p^{2n+1}\rangle$ also approach the same value
of the original MB distribution (zero) {[}Fig.\,\ref{fig-entropy}(c){]}. 

Therefore, in the long time limit, $\mathscr{P}_{\mathtt{p}}(p,t\rightarrow\infty)$
approaches an exotic function discontinuous everywhere, which is different
from the initial MB distribution, but all of its moments $\langle p^{n}\rangle$
have the same values as the initial MB distribution \citep{lin_moment_1997,mayato_m-indeterminate_2018}.
In usual experiments, practically it is difficult to tell the difference
of these two different distributions \citep{swendsen_explaining_2008}. 

\subsection{Correlation entropy \label{Sec:CorrelationEntropy}}

\begin{figure}
\begin{centering}
\includegraphics[width=0.6\textwidth]{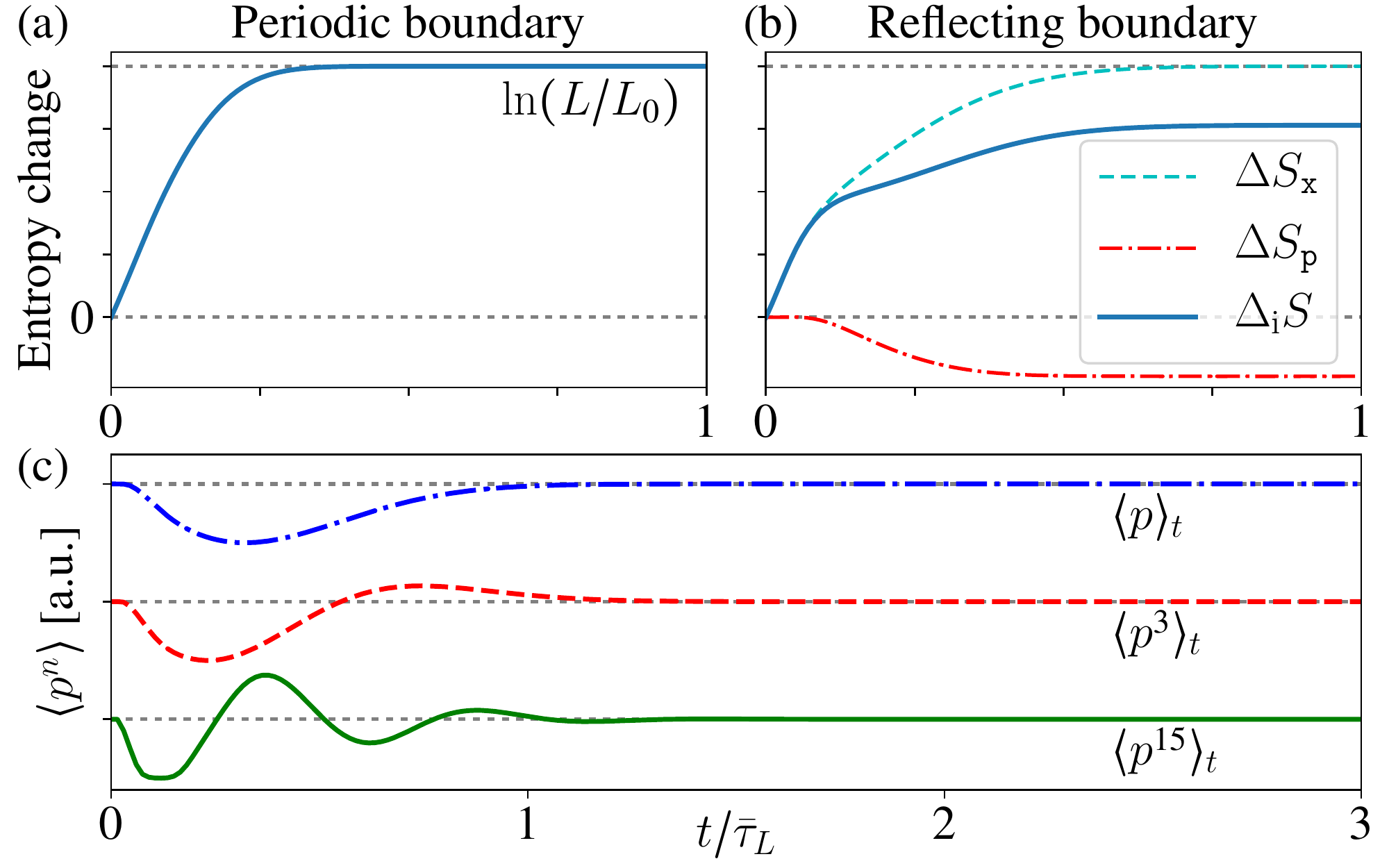}
\par\end{centering}
\caption{The increase of the correlation entropy $\Delta_{\mathrm{i}}S$ for
the (a) periodic and (b) reflecting boundary cases. (c) The evolution
of odd moments $\langle p^{n}\rangle_{t}$ under reflecting boundary
condition (the values are normalized by their maximum amplitudes for
comparison). The unit $\bar{\tau}_{L}:=mL/\bar{p}_{T}$ is the time
for a particle with average kinetic energy $\bar{p}_{T}^{2}/2m$ to
pass $L$.}

\label{fig-entropy}
\end{figure}
From the above exact results for the ensemble evolution, we have seen
that the macroscopic appearance of the new uniform distribution is
indeed only about the spatial distribution, which just reflects the
marginal information of the whole state $\varrho(p,x,t)$, thus this
macroscopic appearance is not enough to conclude whether $\varrho(p,x,t)$
is approaching the new equilibrium state with entropy increase.

However, in practical experiments, the full joint distribution $\varrho(p,x)$
is difficult to be measured directly. Usually it is the spatial and
momentum distributions $\mathscr{P}_{\mathtt{x}}(x)$ and $\mathscr{P}_{\mathtt{p}}(p)$
that are directly accessible for measurements, e.g., by measuring
the gas density and pressure. Therefore, based on these two marginal
distributions, we may ``infer'' the microstate PDF as \citep{jaynes_information_1957,garrett_macroirreversibility_1991,strasberg_quantum_2017}
\begin{equation}
\tilde{\varrho}_{\text{inf}}(p,x,t):=\mathscr{P}_{\mathtt{x}}(x,t)\times\mathscr{P}_{\mathtt{p}}(p,t),
\end{equation}
which indeed neglected the correlation between these two marginal
distributions. As a result, in the long time limit, $\mathscr{P}_{\mathtt{x}}(x,t)\rightarrow1/L$
approaches the new uniform distribution, while $\mathscr{P}_{\mathtt{p}}(x,t)$
``behaves'' similarly like the initial MB distribution, thus this
inferred state $\tilde{\varrho}_{\text{inf}}(p,x,t\rightarrow\infty)$
just looks like a new ``equilibrium state''.

The entropy change of this inferred state is
\begin{align}
\Delta_{\mathrm{i}}S(t):= & S_{\text{\textsc{g}}}[\tilde{\varrho}_{\text{inf}}(t)]-S_{\text{\textsc{g}}}[\tilde{\varrho}_{\text{inf}}(0)]\nonumber \\
= & \Big\{ S_{\mathtt{x}}[\mathscr{P}_{\mathtt{x}}(t)]+S_{\mathtt{p}}[\mathscr{P}_{\mathtt{p}}(t)]-S_{\text{G}}[\varrho(t)]\Big\}-\Big\{ S_{\mathtt{x}}[\mathscr{P}_{\mathtt{x}}(0)]+S_{\mathtt{p}}[\mathscr{P}_{\mathtt{p}}(0)]-S_{\text{G}}[\varrho(0)]\Big\},\label{eq:Entropy}
\end{align}
where $S_{\text{G}}[\varrho(t)]=S_{\text{G}}[\varrho(0)]$ is guaranteed
by the Liouville dynamics, and
\begin{align}
S_{\mathtt{x}}[\mathscr{P}_{\mathtt{x}}(x,t)] & :=-\int_{0}^{L}dx\,\mathscr{P}_{\mathtt{x}}(x,t)\ln\mathscr{P}_{\mathtt{x}}(x,t),\nonumber \\
S_{\mathtt{p}}[\mathscr{P}_{\mathtt{p}}(p,t)] & :=-\int_{-\infty}^{\infty}dp\,\mathscr{P}_{\mathtt{p}}(p,t)\ln\mathscr{P}_{\mathtt{p}}(p,t).\label{eq:Sx Sp}
\end{align}
Notice that the term  $S_{\mathtt{x}}+S_{\mathtt{p}}-S_{\text{\textsc{g}}}:=I_{\mathtt{xp}}$
in Eq.\,(\ref{eq:Entropy}) is just the mutual information between
the marginal distributions $\mathscr{P}_{\mathtt{x}}(x,t)$ and $\mathscr{P}_{\mathtt{p}}(p,t)$,
which is the measure for their correlation \citep{nielsen_quantum_2000,zhou_irreducible_2008}
(see the discussion about the entropy for continuous PDF in Appendix
\ref{sec:PDF entropy}). 

Therefore, here  $\Delta_{\mathrm{i}}S$ just describes the correlation
increase between the spatial and momentum distributions. During the
diffusion process, this correlation entropy $\Delta_{\mathrm{i}}S(t)$
increases monotonically for both periodic and reflecting boundary
cases {[}Fig.\,\ref{fig-entropy}(a, b){]}. Notice that this is quite
similar with the above discussions about open systems, namely, the
total entropy does not change, while the correlation entropy increases
``irreversibly'' \citep{li_production_2017,you_entropy_2018,esposito_entropy_2010,strasberg_quantum_2017,zhang_general_2008,zhang_information_2009,pucci_entropy_2013,horowitz_equivalent_2014,alipour_correlations_2016,manzano_entropy_2016}. 

For the periodic boundary case, $\mathscr{P}_{\mathtt{p}}(p)$ does
not change, and $\mathscr{P}_{\mathtt{x}}(x)$ approaches the uniform
distribution after long time, thus the above entropy increase (\ref{eq:Entropy})
gives $\Delta_{\mathrm{i}}S=\ln(L/L_{0})$, where $L_{0}:=b-a$ is
the length of the initially occupied area. When considering the full
$N$-particle state of the ideal gas, the corresponding inferred state
is $\tilde{\rho}_{\text{inf}}(\vec{P},\vec{Q})=\prod_{i,\sigma}\tilde{\varrho}_{\text{inf}}(p_{i}^{\sigma},q_{i}^{\sigma})$,
thus it  gives the entropy increase as $\Delta_{\mathrm{i}}S=N\ln(V/V_{0})$.
Notice that this result exactly reproduces the above thermodynamic
entropy increase as mentioned in the beginning of this section {[}Fig.\,\ref{fig-entropy}(a){]}
(these conclusions still hold in the thermodynamics limit $V\rightarrow\infty$,
since the system sizes always appear in ratios, e.g., $V/V_{0}$).

For the reflecting boundary case, $\mathscr{P}_{\mathtt{x}}(x)\rightarrow1/L$
still holds and gives $\Delta S_{\mathtt{x}}=\ln(L/L_{0})$, but now
$\mathscr{P}_{\mathtt{p}}(p)$ varies with time. Moreover, it is worthwhile
to notice that $\Delta S_{\mathtt{p}}[\mathscr{P}_{\mathtt{p}}]$
is \emph{decreasing} with time {[}Fig.\,\ref{fig-entropy}(b){]},
which looks a little counter-intuitive. The reason is, as we mentioned
before, the boundary reflections change the momentum directions and
so as the distribution $\mathscr{P}_{\mathtt{p}}(p)$, but the average
energy $\langle p^{2}\rangle$ does not change, thus the thermal distribution
(which is the initial distribution) should have the maximum entropy
\citep{jaynes_information_1957,reichl_modern_2009}. Therefore, during
the evolution, the deviation of $\mathscr{P}_{\mathtt{p}}(p,t)$ from
the initial MB distribution leads to the decrease of its entropy $\Delta S_{\mathtt{p}}[\mathscr{P}_{\mathtt{p}}]$.
However, clearly this is quite difficult to be sensed in practice,
and the total correlation entropy change $\Delta_{\mathrm{i}}S=\Delta S_{\mathtt{x}}+\Delta S_{\mathtt{p}}$
still increases monotonically.

\subsection{Resolution induced coarse-graining \label{Sec:Resolution}}

Historically, the problem of the constant entropy from the Liouville
dynamics was first studied by Gibbs \citep{gibbs_elementary_1902}.
To understand why the entropy increases in the standard thermodynamics,
he noticed that, if we change the order of taking limit when calculating
the ``ensemble volume'' (the entropy), the results are different.
Usually, at a certain time $t$, the phase space is divided into many
small cells with a finite volume $\varepsilon_{V}$, and we make summation
from all these cells, then let the cell size $\varepsilon_{V}\rightarrow0$.
That gives the result of constant entropy. But if we keep the cell
size finite, and let the time $t\rightarrow\infty$ first, then let
the cell size $\varepsilon_{V}\rightarrow0$, that would give an increasing
entropy. 

This idea is now more specifically described as the ``coarse-graining''
\citep{gibbs_elementary_1902,hobson_concepts_1971,uffink_compendium_2006}.
The ``coarse-grained'' ensemble state $\tilde{\rho}_{\text{c.g.}}(\vec{P},\vec{Q})$
is obtained by taking the phase-space average of the exact one $\rho(\vec{P},\vec{Q})$
over a small volume around each point $(\vec{P},\vec{Q})$, namely,
\begin{equation}
\tilde{\rho}_{\text{c.g.}}(\vec{P},\vec{Q}):=\int_{\varepsilon_{V}}d^{3N}P\,d^{3N}Q\,\rho(\vec{P},\vec{Q})\Big/\varepsilon_{V}.
\end{equation}
Here $\varepsilon_{V}$ is the small volume around the point $(\vec{P},\vec{Q})$
in the phase space. From Fig.\,\ref{fig-phasePDF}(d) and Fig.\,\ref{fig-reflect}(d),
we can see after long time relaxation, the coarse-grained $\tilde{\rho}_{\text{c.g.}}(\vec{P},\vec{Q})$
could approach the equilibrium state.

\begin{figure}
\begin{centering}
\includegraphics[width=0.6\textwidth]{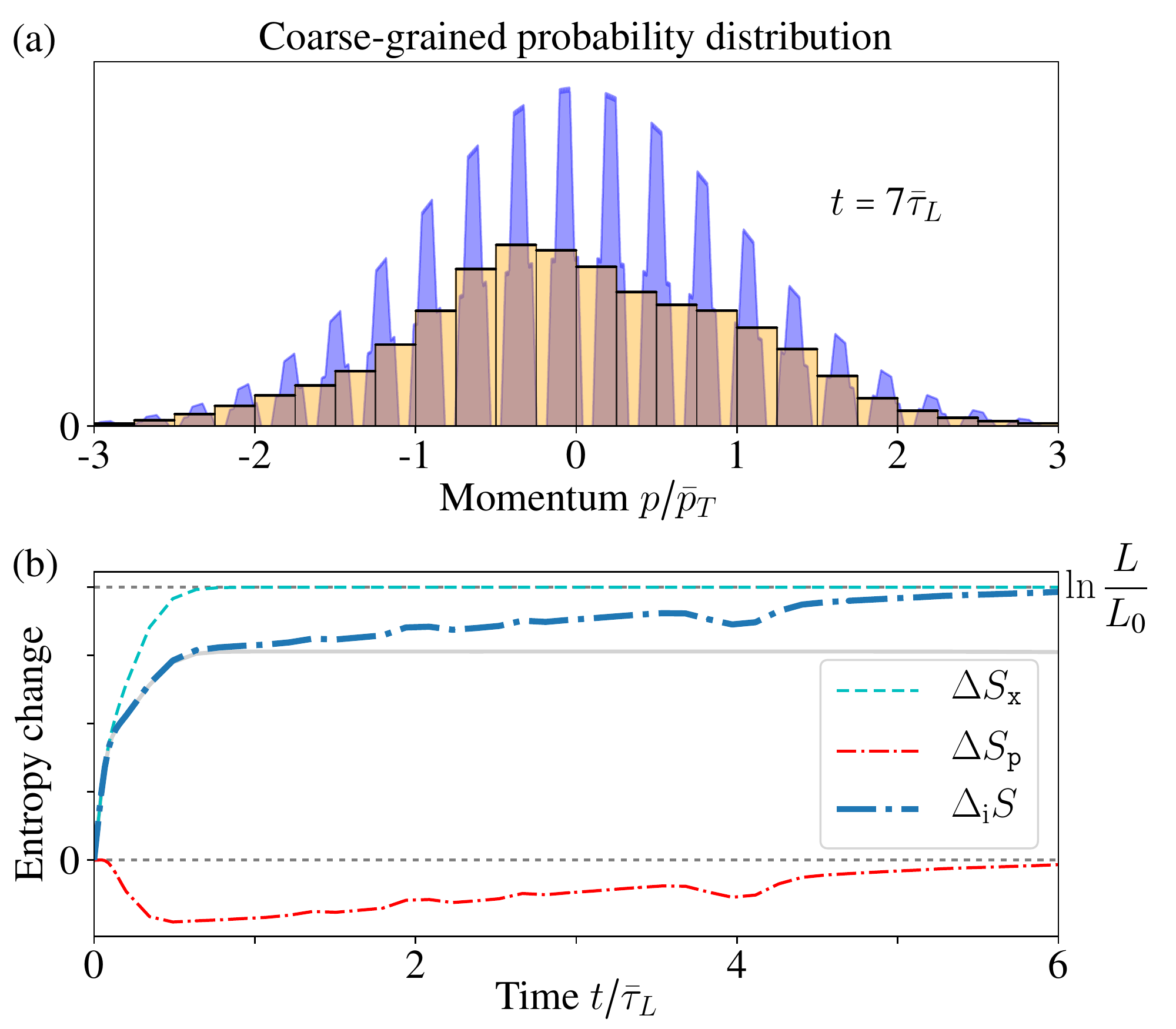}
\par\end{centering}
\caption{(a) The exact distribution $\mathscr{P}_{\mathtt{p}}(p)$ (blue) and
its coarse-grained distribution $\widetilde{\mathscr{P}}_{\mathtt{p}}(p)$
(orange histogram) at $t=7\bar{\tau}_{L}$, which is quite close to
the MB one. (b) The entropy change calculated by the coarse-grained
distribution $\widetilde{\mathscr{P}}_{\mathtt{p}}(p)$, where the
solid gray line is calculated from the exact $\mathscr{P}_{\mathtt{p}}(p)$
for comparison. The parameters are the same with those in Fig.\,\ref{fig-entropy}.}

\label{fig-coarse}
\end{figure}
In practical measurements, there always exists a finite resolution
limit, and that can be regarded as the physical origin of coarse-graining.
When measuring a continuous PDF $P(x)$, we should first divide the
continuous area $x\in[0,L]$ into $N$ intervals, and then measure
the probability that appears in the interval between $x_{n}$ and
$x_{n}+\Delta x$, which is denoted as $p_{n}=P(x_{n})\Delta x$.
In the limit $\Delta x\rightarrow0$, the histogram $P(x_{n})$ becomes
the continuous probability density (see also Appendix \ref{sec:PDF entropy}). 

Here $\Delta x$ is determined by the measurement resolution, but
practical measurements always have a finite resolution limit $\Delta x\gtrsim\delta\tilde{\mathtt{x}}$,
thus cannot approach 0 arbitrarily. Therefore, the fine structure
within the minimum resolution interval $\delta\tilde{\mathtt{x}}$
of the continuous PDF $P(x)$ cannot be sensed in practice. However,
usually $P(x)$ is assumed to be a smooth function within this small
interval, thus indeed $P(x)$ is coarse-grained by its average value
in this small region, and the resolution limit $\delta\tilde{\mathtt{x}}$
practically determines the coarse-graining size.

Remember in the reflecting boundary case, the momentum distribution
$\mathscr{P}_{\mathtt{p}}(p)$ approaches an exotic function with
a structure of dense comb and discontinuous everywhere. Therefore,
such an exotic structure within the resolution limit indeed cannot
be observed in practice. For a fixed measurement resolution $\delta\tilde{p}$,
there always exists a certain time $t_{\delta\tilde{p}}$, so that
after $t>t_{\delta\tilde{p}}$, the ``comb teeth'' in Fig.\,\ref{fig-coarse}(a)
are finer than the resolution $\delta\tilde{p}$. And when $t\gg t_{\delta\tilde{p}}$,
the coarse-grained distribution $\widetilde{\mathscr{P}}_{\mathtt{p}}(p)$
(with coarse-graining size $\delta\tilde{p}$) would well approach
the thermal distribution {[}Fig.\,\ref{fig-coarse}(b){]}. 

In Fig.\,\ref{fig-entropy}(c), we have shown that all the moments
$\langle p^{n}\rangle$ of $\mathscr{P}_{\mathtt{p}}(p,t\rightarrow\infty)$
have no difference with the initial MB distribution. Now due to the
finite resolution limit, again we have no way to tell the difference
between the exotic function $\mathscr{P}_{\mathtt{p}}(p,t\rightarrow\infty)$
and its coarse-graining $\widetilde{\mathscr{P}}_{\mathtt{p}}(p,t\rightarrow\infty)$,
which goes back to the initial MB distribution {[}Fig.\,\ref{fig-coarse}(a){]}.
In this sense, the momentum distribution $\mathscr{P}_{\mathtt{p}}(p,t\rightarrow\infty)$
has no ``practical'' difference with the thermal equilibrium distribution.

\subsection{Entropy ``decreasing'' process \label{sec:Entropy decrease}}

Now we see the correlation entropy, rather than the total entropy,
coincides closer to the irreversible entropy increase in macroscopic
thermodynamics. Here we show it could be possible, although not quite
feasible in practice, to construct an ``entropy decrease'' process. 

To achieve this, we first let the ideal gas experience the above diffusion
process with ``entropy increase'' for a certain time $t_{*}$ (Figs.\,\ref{fig-phasePDF},\,\ref{fig-reflect}).
From the state $\varrho(p,x,t_{*})$ at this moment {[}e.g., Fig.\,\ref{fig-reflect}(d){]},
we construct a new ``initial state'' by reversing its momentum $\varrho'(0)=\varrho(-p,x,t_{*})$.
Since the Liouville equation obeys time-reversal symmetry, this new
``initial state'' would evolve into $\varrho'(t_{*})=\varrho(-p,x,0)$
after time $t_{*}$, which is just the original equilibrium state
confined in $x\in[a,b]$ {[}Fig.\,\ref{fig-reflect}(a){]}. That
means, the idea gas exhibits a process of ``reversed diffusion''.

During this time-reversal process, the total Gibbs entropy, which
contains the full information, still keeps constant. However, the
correlation entropy change $\Delta_{\text{i}}S$ (no matter whether
coarse-grained) would exactly experience the reversed ``backward''
evolution of Fig.\,\ref{fig-entropy}, which is an ``entropy decreasing''
process.

This is just the idea of the Loschmidt paradox \citep{cohen_history_2005,uffink_compendium_2006,zaslavsky_chaotic_2008,lebowitz_boltzmanns_2008,brown_boltzmanns_2008},
except two subtle differences: 1. here we are talking about the ideal
gas with no particle collision, but the original Loschmidt paradox
was about the Boltzmann transport equation in the presence of particle
collisions; 2. here we are talking about the correlation entropy between
the spatial and momentum distributions, while the Boltzmann equation
is about the entropy of the single-particle PDF.

We must notice that such an initial state $\varrho'(0)$ is NOT an
equilibrium state, but contains very delicate correlations between
the spatial and momentum distributions \citep{jaynes_gibbs_1965}.
In order to see such a time-reversal process, the initial state must
be precisely prepared to contain such specific correlations between
the marginal distributions {[}Fig.\,\ref{fig-reflect}(d){]}, which
is definitely quite difficult for practical operation. Therefore,
such an ``entropy decrease'' process is rarely seen in practice
(except some special cases like the Hahn echo \citep{hahn_spin_1950}
and back-propagating wave \citep{fink_time_2008,przadka_time_2012}).

\section{The correlation in the Boltzmann equation }

In the above section, we focused on the diffusion of the ideal gas
with no inter-particle interactions, and the initial momentum distribution
has been assumed to be the MB distribution in priori. If the initial
momentum distribution is not the MB one, the particles could still
exhibit the irreversible diffusion filling the whole volume, but $\mathscr{P}_{\mathtt{p}}(p,t)$
would never become the thermal equilibrium distribution.

If there exist weak interactions between the particles, as shown by
the Boltzmann \emph{H-}theorem, the single-particle PDF $f(\mathbf{p},\mathbf{r},t)$
could always approach the MB distribution as its steady state, together
with the irreversible entropy increase. And in this case, the above
``coarse-graining'' is not needed for $f(\mathbf{p},\mathbf{r},t)$
to approach the thermal equilibrium distribution.

Notice that, in the Boltzmann \emph{H-}theorem, only the single-particle
PDF $f(\mathbf{p},\mathbf{r},t)$ is concerned. The total ensemble
$\rho(\vec{P},\vec{Q},t)$ in the $6N$-dimensional phase space should
still follow the Liouville equation, thus its entropy does not change
with time. Therefore, the \emph{H}-theorem conclusion also can be
understood as the inter-particle correlations are increasing irreversibly
\citep{jaynes_gibbs_1965,huang_statistical_1987,kalogeropoulos_time_2018}.
Again, this is quite similar like the correlation understanding in
the last two sections, and here the inter-particle correlation entropy
could well reproduce the entropy increase result in the standard macroscopic
thermodynamics.

The debates about the Boltzmann \emph{H}-theorem started ever since
its birth. The most important one must be the Loschmidt paradox raised
in 1876: due to the time-reversal symmetry of the microscopic dynamics
of the particles, once their momentums are reversed at the same time,
the particles should follow their incoming paths ``backward'', which
is surely a possible evolution for the microstate; however, if the
entropy of $f(\mathbf{p},\mathbf{r},t)$ must increase (according
to the \emph{H}-theorem), then the corresponding ``backward'' evolution
must give an entropy decreasing process, which is contradicted with
the \emph{H}-theorem conclusion.

In this section, we will see the slowly increasing inter-particle
correlations could be helpful in understanding this paradox. Namely,
due to the significant inter-particle correlation established during
the ``forward'' process, indeed the ``backward'' process no longer
satisfies the molecular-disorder assumption, which is a crucial approximation
in deriving the Boltzmann equation, thus it is not suitable to be
described by the Boltzmann equation, and the \emph{H}-theorem does
not apply in this case either.

\subsection{Derivation of the Boltzmann equation }

We first briefly review the derivation of the Boltzmann transport
equation \citep{boltzmann_vorlesungen_1896,boltzmann_lectures_1964,huang_statistical_1987}.
When there is no external force, the evolution equation of the single-particle
microstate PDF $f(\mathbf{p}_{1},\mathbf{r}_{1},t)$ is 
\begin{equation}
\Big[\partial_{t}+\frac{\mathbf{p}_{1}}{m}\cdot\nabla_{\mathbf{r}_{1}}\Big]f(\mathbf{p}_{1},\mathbf{r}_{1},t)=\partial_{t}f\big|_{\text{col}}.\label{eq:BE-0}
\end{equation}
 The left side is just the above Liouville equation (\ref{eq:Liouville})
of the ideal gas, and the right side is the probability change due
to the particle collision (assuming only bipartite collisions exist). 

This collision term, rewritten as $\partial_{t}f\big|_{\text{col}}=\Delta^{(+)}-\Delta^{(-)}$,
contains two contributions: $\Delta^{(-)}$ means the collision between
two particles $(\mathbf{p}_{1}\mathbf{r}_{1};\,\mathbf{p}_{2}\mathbf{r}_{2})\rightarrow(\mathbf{p}_{1}'\mathbf{r}_{1}';\,\mathbf{p}_{2}'\mathbf{r}_{2}')$
kicks particle-1 out of its original region around $(\mathbf{p}_{1},\mathbf{r}_{1})$,
thus $f(\mathbf{p}_{1},\mathbf{r}_{1},t)$ decreases; likewise, $\Delta^{(+)}$
means the collision $(\mathbf{p}_{1}'\mathbf{r}_{1}';\,\mathbf{p}_{2}'\mathbf{r}_{2}')\rightarrow(\mathbf{p}_{1}\mathbf{r}_{1};\,\mathbf{p}_{2}\mathbf{r}_{2})$
kicks particle-1 into the region around $(\mathbf{p}_{1},\mathbf{r}_{1})$
and that increases $f(\mathbf{p}_{1},\mathbf{r}_{1},t)$. 

\begin{figure}
\begin{centering}
\includegraphics[width=0.7\textwidth]{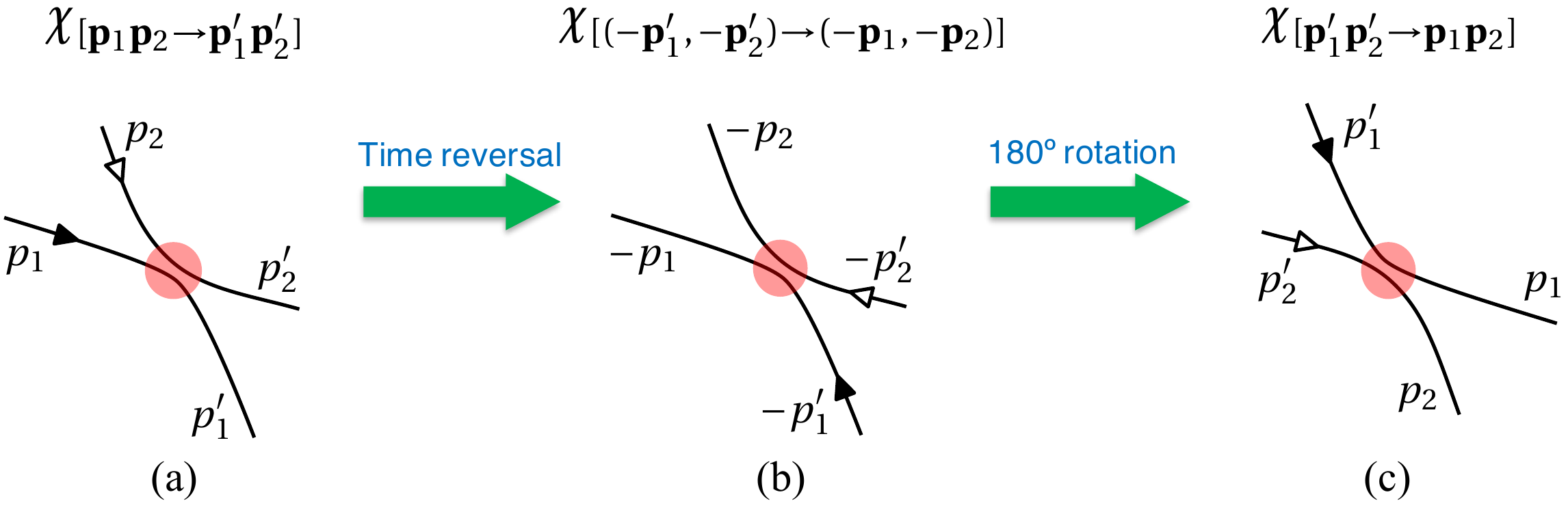}
\par\end{centering}
\caption{Assume collisions happen only in short range, then the transition
rate $\chi_{[(\mathbf{p}_{1}\mathbf{r}_{1};\mathbf{p}_{2}\mathbf{r}_{2})\rightarrow(\mathbf{p}_{1}'\mathbf{r}_{1}';\mathbf{p}_{2}'\mathbf{r}_{2}')]}$
could have nonzero value only within the range $\mathbf{r}_{1}\simeq\mathbf{r}_{2}\simeq\mathbf{r}_{1}'\simeq\mathbf{r}_{2}'$.
The scattering process (b) is the time reversal of (a), thus they
have equal transition rates $\chi_{[\mathbf{p}_{1}\mathbf{p}_{2}\rightarrow\mathbf{p}_{1}'\mathbf{p}_{2}']}=\chi_{[(-\mathbf{p}_{1}',-\mathbf{p}_{2}')\rightarrow(-\mathbf{p}_{1},-\mathbf{p}_{2})]}$.
The scattering process (c) is obtained by making the $180^{\circ}$
inversion of (b), thus they also have equal transition rates $\chi_{[(-\mathbf{p}_{1}',-\mathbf{p}_{2}')\rightarrow(-\mathbf{p}_{1},-\mathbf{p}_{2})]}=\chi_{[\mathbf{p}_{1}'\mathbf{p}_{2}'\rightarrow\mathbf{p}_{1}\mathbf{p}_{2}]}$.
Therefore, we obtain the relation $\chi_{[12\rightarrow1'2']}=\chi_{[1'2'\rightarrow12]}$.}

\label{fig-scatter}
\end{figure}
These two collision contributions can be further written down as (denoting
$\mathrm{d}\boldsymbol{\varsigma}_{i}:=d^{3}\mathbf{r}_{i}\,d^{3}\mathbf{p}_{i}$)
\begin{align}
\Delta^{(-)} & =\int F(\mathbf{p}_{1}\mathbf{r}_{1};\,\mathbf{p}_{2}\mathbf{r}_{2},t)\chi_{[12\rightarrow1'2']}\,\mathrm{d}\boldsymbol{\varsigma}_{1}'\,\mathrm{d}\boldsymbol{\varsigma}_{2}'\,\mathrm{d}\boldsymbol{\varsigma}_{2},\nonumber \\
\Delta^{(+)} & =\int F(\mathbf{p}_{1}'\mathbf{r}_{1}';\,\mathbf{p}_{2}'\mathbf{r}_{2}',t)\chi_{[1'2'\rightarrow12]}\,\mathrm{d}\boldsymbol{\varsigma}_{1}'\,\mathrm{d}\boldsymbol{\varsigma}_{2}'\,\mathrm{d}\boldsymbol{\varsigma}_{2},
\end{align}
where $F(\mathbf{p}_{1}\mathbf{r}_{1};\,\mathbf{p}_{2}\mathbf{r}_{2},t)$
is the two-particle joint probability, and $\chi_{[12\rightarrow1'2']}$
denotes the transition ratio (or scattering matrix) from the initial
state $(\mathbf{p}_{1}\mathbf{r}_{1};\,\mathbf{p}_{2}\mathbf{r}_{2})$
scattered into the final state $(\mathbf{p}_{1}'\mathbf{r}_{1}';\,\mathbf{p}_{2}'\mathbf{r}_{2}')$.
Due to the time-reversal and inversion symmetry of the microscopic
scattering process, the transition ratios $\chi_{[12\rightarrow1'2']}$
and $\chi_{[1'2'\rightarrow12]}$ equal to each other (see Fig.\,\ref{fig-scatter}).
Therefore, the above equation\,(\ref{eq:BE-0}) is further written
as {[}denoting $F_{12}:=F(\mathbf{p}_{1}\mathbf{r}_{1};\,\mathbf{p}_{2}\mathbf{r}_{2},t)${]}
\begin{equation}
\partial_{t}f(\mathbf{p}_{1},\mathbf{r}_{1},t)+\frac{\mathbf{p}_{1}}{m}\cdot\nabla_{\mathbf{r}_{1}}f=\int(F_{1'2'}-F_{12})\chi_{[12\rightarrow1'2']}\,\mathrm{d}\boldsymbol{\varsigma}_{1}'\,\mathrm{d}\boldsymbol{\varsigma}_{2}'\,\mathrm{d}\boldsymbol{\varsigma}_{2}.\label{eq:BE-1}
\end{equation}

Now we adopt the ``\emph{molecular-disorder assumption}'', i.e.,
the two-particle joint PDF can be approximately written as the product
of the two single-particle PDF\footnote{Here we adopt the wording
from Boltzmann's original paper \citep{boltzmann_vorlesungen_1896}
(English translation \citep{boltzmann_lectures_1964}). In literature
this assumption is usually called \emph{Stosszahlansatz}, or the \emph{molecular
chaos hypothesis}. The word ``\emph{Stosszahlansatz}'' was first
used by Ehrenfest in 1912, and its original meaning is ``the assumption
of collision number'' \citep{ehrenfest_conceptual_2015,uffink_compendium_2006,brown_boltzmanns_2008}.
}
\begin{align}
F(\mathbf{p}_{1}\mathbf{r}_{1};\mathbf{p}_{2}\mathbf{r}_{2},t) & \simeq f(\mathbf{p}_{1},\mathbf{r}_{1},t)\times f(\mathbf{p}_{2},\mathbf{r}_{2},t).\label{eq:MoleDisorder}
\end{align}
Essentially this is requiring that the correlation between the two
particles is negligible. Then we obtain the Boltzmann transport equation
{[}denoting $f_{i}:=f(\mathbf{p}_{i},\mathbf{r}_{i},t)${]}
\begin{equation}
\partial_{t}f(\mathbf{p}_{1},\mathbf{r}_{1},t)+\frac{\mathbf{p}_{1}}{m}\cdot\nabla_{\mathbf{r}_{1}}f=\int(f_{1'}f_{2'}-f_{1}f_{2})\chi_{[12\rightarrow1'2']}\,\mathrm{d}\boldsymbol{\varsigma}_{1}'\,\mathrm{d}\boldsymbol{\varsigma}_{2}'\,\mathrm{d}\boldsymbol{\varsigma}_{2}.\label{eq:Boltzmann}
\end{equation}

\subsection{\emph{H}-theorem and the steady state }

Now we further review how to prove the \emph{H}-theorem from the above
Boltzmann equation, and find out its steady state. Defining the Boltzmann
\emph{H}-function as $H[f(\mathbf{p}_{1},\mathbf{r}_{1},t)]:=\int\mathrm{d}\boldsymbol{\varsigma}_{1}\,f_{1}\ln f_{1}$,
the Boltzmann equation\,(\ref{eq:Boltzmann}) guarantees $H(t)$
decreases monotonically ($dH/dt\le0$), and this is the \emph{H}-theorem.

To prove this theorem, we put the Boltzmann equation into the time
derivative $\frac{d}{dt}H(t)=\int\mathrm{d}\boldsymbol{\varsigma}_{1}\,\partial_{t}f(\mathbf{p}_{1},\mathbf{r}_{1},t)\cdot\ln f(\mathbf{p}_{1},\mathbf{r}_{1},t)$.
The Liouville diffusion term gives (denoting $\mathbf{v}:=\mathbf{p}/m$)
\begin{equation}
\int\mathrm{d}\boldsymbol{\varsigma}_{1}\,\frac{\mathbf{p}_{1}}{m}\cdot\nabla_{\mathbf{r}_{1}}f_{1}\cdot\ln f_{1}=\int\mathrm{d}\boldsymbol{\varsigma}\,\nabla_{\mathbf{r}}\cdot(\mathbf{v}f\ln f-\mathbf{v}f),
\end{equation}
 which can be turned into a surface integral and vanishes. And the
collision term gives
\begin{align}
\frac{dH}{dt}= & \int(f_{1'}f_{2'}-f_{1}f_{2})\chi_{[12\rightarrow1'2']}\ln f_{1}\,\mathrm{d}\boldsymbol{\varsigma}_{1}'\mathrm{d}\boldsymbol{\varsigma}_{2}'\mathrm{d}\boldsymbol{\varsigma}_{1}\mathrm{d}\boldsymbol{\varsigma}_{2}\nonumber \\
= & \frac{1}{2}\int(f_{1'}f_{2'}-f_{1}f_{2})\chi_{[12\rightarrow1'2']}\ln f_{1}\,\mathrm{d}\boldsymbol{\varsigma}_{1}'\mathrm{d}\boldsymbol{\varsigma}_{2}'\mathrm{d}\boldsymbol{\varsigma}_{1}\mathrm{d}\boldsymbol{\varsigma}_{2}+\frac{1}{2}\int(f_{2'}f_{1'}-f_{2}f_{1})\chi_{[21\rightarrow2'1']}\ln f_{2}\,\mathrm{d}\boldsymbol{\varsigma}_{2}'\mathrm{d}\boldsymbol{\varsigma}_{1}'\mathrm{d}\boldsymbol{\varsigma}_{2}\mathrm{d}\boldsymbol{\varsigma}_{1}\nonumber \\
= & \frac{1}{2}\int\left[(f_{1'}f_{2'}-f_{1}f_{2})\chi_{[12\rightarrow1'2']}\ln(f_{1}f_{2})\right]\mathrm{d}\boldsymbol{\varsigma}_{1}'\mathrm{d}\boldsymbol{\varsigma}_{2}'\mathrm{d}\boldsymbol{\varsigma}_{1}\mathrm{d}\boldsymbol{\varsigma}_{2},
\end{align}
where the second line is because exchanging the integral variables
$1\leftrightarrow2$ gives the same value. We can further apply the
similar trick by exchanging the integral variables $(12)\leftrightarrow(1'2')$,
and that gives 
\begin{equation}
\frac{dH}{dt}=\frac{1}{4}\int(f_{1'}f_{2'}-f_{1}f_{2})(\ln f_{1}f_{2}-\ln f_{1'}f_{2'})\chi_{[12\rightarrow1'2']}\,\mathrm{d}\boldsymbol{\varsigma}_{1}'\mathrm{d}\boldsymbol{\varsigma}_{2}'\mathrm{d}\boldsymbol{\varsigma}_{1}\mathrm{d}\boldsymbol{\varsigma}_{2}.
\end{equation}
Here the transition ratio $\chi_{[12\rightarrow1'2']}$ is non-negative,
and notice that $(f_{1'}f_{2'}-f_{1}f_{2})(\ln f_{1}f_{2}-\ln f_{1'}f_{2'})\le0$
always holds for any PDF $f_{i}$. Therefore, we obtain $dH/dt\le0$,
which means the function $H(t)$ decreases monotonically, and this
encloses the proof. $\;\blacksquare$

In the above inequality, the equality holds if and only if $f_{1}f_{2}=f_{1'}f_{2'}$,
which means the collision induced increase $\Delta^{(+)}$ and decrease
$\Delta^{(-)}$ of $f(\mathbf{p},\mathbf{r})$ must balance each other
everywhere, thus it is also known as the detailed balance condition. 

The time-independent steady state of $f(\mathbf{p},\mathbf{r},t)$
can be obtained from this detailed balance equation $f_{1}f_{2}=f_{1'}f_{2'}$.
Taking the logarithm of the two sides, it gives
\begin{equation}
\ln f(\mathbf{p}_{1},\mathbf{r}_{1})+\ln f(\mathbf{p}_{2},\mathbf{r}_{2})=\ln f(\mathbf{p}_{1}',\mathbf{r}_{1}')+\ln f(\mathbf{p}_{2}',\mathbf{r}_{2}').
\end{equation}
Notice that the two sides of the above equation depends on different
variables, and has a conservation form. Therefore, $\ln f$ must be
a combination of some conservative quantities. During the collision
$(\mathbf{p}_{1}\mathbf{r}_{1};\,\mathbf{p}_{2}\mathbf{r}_{2})\leftrightarrow(\mathbf{p}_{1}'\mathbf{r}_{1}';\,\mathbf{p}_{2}'\mathbf{r}_{2}')$,
the particles collides at the same position, and the total momentum
and energy are conserved, thus $\ln f$ must be their combinations,
namely, $\ln f=C_{0}+\mathbf{C}_{1}\cdot\mathbf{p}+C_{2}\mathbf{p}^{2}$,
where $C_{0}$, $\mathbf{C}_{1}$, $C_{2}$ are constants. Therefore,
$f(\mathbf{p},\mathbf{r})$ must be a Gaussian distribution of $\mathbf{p}$
at any position $\mathbf{r}$. 

Further, the diffusion term in Eq.\,(\ref{eq:Boltzmann}) requires
$\mathbf{p}\cdot\nabla_{\mathbf{r}}f=0$ in the steady state, thus
$f(\mathbf{p},\mathbf{r})$ must be homogenous for any position $\mathbf{r}$.
The average momentum $\langle\mathbf{p}\rangle$ should be 0 for a
stationary gas. Therefore, the steady state of the Boltzmann equation\,(\ref{eq:Boltzmann})
is a Gaussian distribution  $f(\mathbf{p},\mathbf{r})\sim\exp[-\mathbf{p}^{2}/2\overline{p}_{T}^{2}]$
independent of the position $\mathbf{r}$, which is the MB distribution.

\subsection{Molecular-disorder assumption and Loschmidt paradox }

In the above two sections, we demonstrated all the critical steps
deriving the Boltzmann equation. Notice that there is no special requirement
for the interaction form of the collisions, as long as it is short-ranged
so as to make sure only bipartite collisions exist. The contribution
of the collision interaction is implicitly contained in the transition
rate $\chi_{[12\rightarrow1'2']}$, and the only properties we utilized
are (1) $\chi_{[12\rightarrow1'2']}\ge0$ and (2) $\chi_{[12\rightarrow1'2']}=\chi_{[1'2'\rightarrow12]}$.
Thus it does not matter whether the interaction is nonlinear.

No doubt to say, the molecular-disorder assumption ($F_{12}\simeq f_{1}\times f_{2}$)
is the most important basis in the above derivations\footnote{ In
the last paragraph of Chap. I-5, Part I of Ref.\,\citep{boltzmann_lectures_1964}
(pp. 29), Boltzmann said, ``\emph{...The only assumption made here
is that the velocity distribution is molecular-disordered }(namely,
$F_{12}\simeq f_{1}\times f_{2}$ in our notation)\emph{ at the beginning,
and remains so. With this assumption, one can prove that H can only
decrease, and also that the velocity distribution must approach that
of Maxwell.}'' }. Before this approximation, indeed Eq.\,(\ref{eq:BE-1})
is still formally exact. Clearly, the validity of this assumption,
which is imposed on the particle correlations, determines whether
the Boltzmann equation (\ref{eq:Boltzmann}) holds. Now we will re-examine
this assumption as well as the Loschmidt paradox.

Once two particles collide with each other, they get correlated. In
a dilute gas, collisions do not happen very frequently, and once two
particles collides with each other, they could hardly meet each other
again. Therefore, if initially there is no correlations between particles,
we can expect that, on average, the collision induced bipartite correlations
are negligibly small, and thus this molecular-disorder assumption
holds well. 

Now we look at the situation in the Loschmidt paradox. First, the
particles experience a ``forward'' diffusion process for a certain
time. According to the \emph{H}-theorem, the entropy increases in
this process. Then suppose all the particle momentums are suddenly
reversed at this moment. From this initial state, the particles are
supposed to evolve ``backward'' exactly along the incoming trajectory,
and thus exhibit an entropy decreasing process, which is contradicted
with the \emph{H}-theorem conclusion, and this is the Loschmidt paradox.

However, we should notice that, indeed the first ``forward'' evolution
has established significant (although very small in quantity) bipartite
correlations in this new ``initial state'' \citep{chliamovitch_kinetic_2017}.
This is quite similar like the above discussion about the momentum-position
correlation in Sec.\,\ref{sec:Entropy decrease}. Thus, the above
molecular-disorder assumption $F_{12}\simeq f_{1}\times f_{2}$ does
not apply in this case. As a result, the next ``backward'' evolution
is indeed unsuitable to be described by Boltzmann equation (\ref{eq:Boltzmann}).
Therefore, the entropy increasing conclusion of the \emph{H}-theorem
($dH/dt\le0$) does not need to hold for this ``backward'' process.
As well, the preparation of such a specific initial state is definitely
unfeasible in practice, thus the ``backward'' entropy decreasing
process is rarely seen.

We emphasize that the Boltzmann equation (\ref{eq:Boltzmann}) is
about the single-particle PDF $f(\mathbf{p},\mathbf{r},t)$, which
is obtained by averaging over all the other $N-1$ particles from
the full ensemble $\rho(\vec{P},\vec{Q},t)$. Clearly, $f(\mathbf{p},\mathbf{r},t)$
omits much information in $\rho(\vec{P},\vec{Q},t)$, but indeed it
is enough to give most macroscopic thermodynamic quantities. For example,
the average kinetic energy of each single molecular $\langle\mathbf{p}^{2}\rangle=\int d\boldsymbol{\varsigma}\,\mathbf{p}^{2}f(\mathbf{p},\mathbf{r})$
determines the gas temperature $T$, and the gas pressure on the wall
is given by $P=\int_{p_{x}>0}d\boldsymbol{\varsigma}\,(2p_{x})\cdot v_{x}f(\mathbf{p},\mathbf{r})$
\citep{huang_statistical_1987}.

In contrast, indeed the inter-particle correlations ignored by the
single-particle PDF $f(\mathbf{p},\mathbf{r},t)$ are quite difficult
to be sensed in practice. Therefore, the $N$-particle ensemble may
be ``inferred'' as $\tilde{\rho}_{\text{inf}}(\vec{P},\vec{Q},t)=\prod_{i=1}^{N}f(\mathbf{p}_{i},\mathbf{r}_{i},t)$,
which clearly omits the inter-particle correlations in the exact $\rho(\vec{P},\vec{Q},t)$.
Similar like the discussion in Sec.\,\ref{Sec:CorrelationEntropy},
based on this inferred ensemble $\tilde{\rho}_{\text{inf}}(\vec{P},\vec{Q},t)$,
the entropy change gives $\Delta_{\mathrm{i}}S=S_{\text{\textsc{g}}}[\tilde{\rho}_{\text{inf}}(t)]-S_{\text{\textsc{g}}}[\tilde{\rho}_{\text{inf}}(0)]=N\ln(V/V_{0})$,
which exactly reproduces the result in the standard thermodynamics.
Thus $\Delta_{\mathrm{i}}S$ indeed characterizes the increase of
the inter-particle correlations.

In sum, even in the presence of the particle collisions, the full
$N$-particle ensemble $\rho(\vec{P},\vec{Q},t)$ still follows the
Liouville equation exactly, thus its Gibbs entropy does not change.
On the other hand, the single-particle PDF $f(\mathbf{p},\mathbf{r},t)$
follows the Boltzmann equation, thus its entropy keeps increasing
until reaching the steady state. And this roots from our ignorance
of the inter-particle correlations in the full $\rho(\vec{P},\vec{Q},t)$. 

\section{Summary}

In this paper, we study the correlation production in open and isolated
thermodynamic systems. In a many-body system, the microscopic dynamics
of the whole system obeys the time-reversal symmetry, which guarantees
the entropy of the global state does not change with time. Based on
the microscopic dynamics, indeed the full ensemble state is not evolving
towards the new equilibrium state as expected from the macroscopic
intuition. However, the correlation between different local DoF, as
measured by their mutual information, generally increases monotonically,
and its amount could well reproduce the entropy increase result in
the standard macroscopic thermodynamics. 

In open systems, as described by the second law in the standard thermodynamics,
the irreversible entropy production increases monotonically. It turns
out that this irreversible entropy production is just equal to the
correlation production between the system and its environment. Thus,
the second law can be equivalently understood as the system-bath correlation
is increasing monotonically, while at the same time, the \noun{s+b
}system as a whole still keeps constant entropy.

In isolated systems, there is no specific partition for ``system''
and ``environment'', but we could see the momentum and spatial distributions,
as the marginal distributions of the total ensemble, exhibit the macroscopic
irreversibility, and their correlation increases monotonically, which
reproduces the entropy increase result in the standard thermodynamics.
In the presence of particle collisions, different particles are also
establishing correlations between each other. As a result, the single-particle
distribution exhibits the macroscopic irreversibility as well as the
entropy increase in the standard thermodynamics, which is just the
result of the Boltzmann \emph{H}-theorem. At the same time, the full
ensemble $\rho(\vec{P},\vec{Q},t)$ of the many-body system still
follows the Liouville equation, which guarantees its entropy does
not change with time.

It is worth noticing that, in practice, usually it is the partial
information (e.g., marginal distribution, few-body observable expectations)
that is directly accessible to our observation. But indeed most macroscopic
thermodynamic quantities are obtained only from such partial information
like the one-body distribution, and that is why they exhibits irreversible
behaviors. However, due to the practical restrictions in measurements,
the dynamics of the full ensemble state, such as the constant entropy
behavior, is quite difficult to be sensed in practice.

In sum, the global state keeps constant entropy, while partial information
exhibits the irreversible entropy increase. But in practice, it is
the partial information that is directly observed. In this sense,
the macroscopic irreversible entropy increase does not contradict
with the microscopic reversibility. Clearly, such correlation production
understanding can be applied for both quantum and classical systems,
no matter whether there exist complicated particle interactions, and
it can be well used for time-dependent non-equilibrium states. Moreover,
it is worths noticing that, if the bath of an open system is a non-thermal
state, indeed this is beyond the application scope of the standard
thermodynamics, but we could see such correlation production understanding
still applies in this case. We notice that such correlations can be
found in many of recent studies of thermodynamics, and it is also
quite interesting to notice that similar idea can be used to understand
the paradox of blackhole information loss, where the mutual information
of the radiation particles is carefully considered \citep{zhang_hidden_2009,zhang_entropy_2011,ma_dark_2018,ma_non-thermal_2018}.

\vspace{1em}

\emph{Acknowledgement} -- SWL appreciates very much for the helpful
discussions with R. Brick, M. B. Kim, R. Nessler, M. O. Scully, A.
Svidzinsky, Z. Yi, L. Zhang in Texas A\&M University, L. Cohen in
City University of New York, and H. Dong in Chinese Academy of Engineering
Physics. This study is supported by the Beijing Institute of Technology
Research Fund Program for Young Scholars.

\appendix

\section{The entropy of a continuous probability distribution \label{sec:PDF entropy}}

For a finite sample space of $N$ events with probabilities $\{p_{n}\}$,
the information entropy is well-defined as $S\big[\{p_{n}\}\big]:=-\sum_{n}p_{n}\ln p_{n}$.
However, the situation of a continuous PDF is not so trivial, and
a simple generalization from the discrete case would lead to a divergency
problem. 

For example, considering a uniform PDF $P(x)=1/L$ in the area $x\in[0,L]$,
to calculate its entropy, we first divide the area into $N$ pieces
averagely, and then turn the discrete summation into the continuous
integral by taking the limit $N\rightarrow\infty$. Clearly here each
piece takes the probability $p_{n}=1/N$, thus the entropy is $S=-\sum_{n}\frac{1}{N}\ln\frac{1}{N}=\ln N$,
but it diverges when $N\rightarrow\infty$. 

Generally, for a continuous PDF $P(x)$ in the area $x\in[0,L]$,
after the division into $N$ pieces, the entropy is 
\begin{equation}
S^{(N)}=-\sum_{n=0}^{N-1}P(x_{n})\Delta x\cdot\ln\big[P(x_{n})\Delta x\big],
\end{equation}
where $\Delta x=L/N$, and $x_{n}=n\cdot\Delta x$. Remember here
$P(x)$ is the probability density, which has the unit of the length
inverse $[L^{-1}]$, and $P(x)\Delta x$ is the unitless probability.
However, when taking the limit $N\rightarrow\infty$, the above equation
becomes
\begin{align}
\lim_{N\rightarrow\infty}S^{(N)} & =\lim_{N\rightarrow\infty}\sum_{n=0}^{N-1}-P(x_{n})\ln P(x_{n})\cdot\Delta x-P(x_{n})\cdot\Delta x\ln\frac{L}{N}\nonumber \\
 & =-\int_{0}^{L}dx\,P(x)\ln P(x)-\int_{0}^{L}dx\,P(x)\ln L+\lim_{N\rightarrow\infty}\ln N\cdot\sum_{n=0}^{N-1}P(x_{n})\Delta x\nonumber \\
 & :=-\int_{0}^{L}dx\,P(x)\ln[P(x)\cdot L]+\ln\mathfrak{N}.
\end{align}
In the 3rd term of the 2nd line, the summation converges to 1 when
$N\rightarrow\infty$, thus this term diverges as $\sim\ln N$, and
we denote it as $\ln\mathfrak{N}$. And notice that in the 1st term,
now $[P(x)\cdot L]$ appears in the logarithm as a whole unitless
quantity.

The diverging term $\ln\mathfrak{N}$ cannot be simply omitted. Considering
the above example of the uniform distribution $P(x)=1/L$ in the area
$x\in[0,L]$, which is supposed to give the largest entropy, we can
see the 1st term in the above result gives 0, and its entropy is exactly
given by this diverging term $S^{(N)}=\ln N$.

Therefore, when generalizing the information entropy for the continuous
PDF, it always contains a diverging term $\ln\mathfrak{N}$. There
are two ways to resolve this problem. First, whenever this entropy
is under discussion, we always focus on the entropy difference between
two states but not their absolute values. For example, in Sec.\,\ref{Sec:CorrelationEntropy}
we always focus on the entropy change $\Delta S:=S(t)-S(0)$ comparing
with the initial state. In this case, the divergency of $\ln\mathfrak{N}$
in $S(t)$ and $S(0)$ just cancels each other. As well, the length
$L$ in the above $\ln[P(x)\cdot L]$ also can be canceled. Therefore,
in the definition\,(\ref{eq:Sx Sp}) for $S_{\mathtt{x}}$ and $S_{\mathtt{p}}$,
the probability densities $\mathscr{P}_{\mathtt{x}}(x)$ and $\mathscr{P}_{\mathtt{p}}(p)$
appear in the logarithm directly although they are not unitless.

Second, from the above simple example of the uniform PDF, we can see
indeed this divergency origins from the idealization for the continuity
of the probability distribution. Notice that a continuous probability
density is indeed not directly accessible in practical measurements,
in contrast, we should first divide the continuous area $x\in[0,L]$
into $N$ intervals, and then measure the probability that appears
in the interval between $x_{n}$ and $x_{n}+\Delta x$, which is denoted
as $p_{n}=P(x_{n})\Delta x$. In the limit $\Delta x\rightarrow0$,
the histogram $P(x_{n})$ becomes the continuous probability density
(see Sec.\,\ref{Sec:Resolution}). Indeed here $\Delta x$ is the
resolution in the measurement. A finer resolution indicates a larger
sample space and more possible probability distributions, and that
results to the divergence of $\ln\mathfrak{N}$. However, most practical
measurements have certain resolution limit, thus $\Delta x$ cannot
approach 0 infinitely. If this resolution restriction is considered,
the diverging term $\ln\mathfrak{N}$ is constrained by the finite
resolution.

\bibliographystyle{apsrev4-1}
\bibliography{Refs}

\end{document}